\documentclass[useAMS,usenatbib]{mn2e}
\usepackage{times}
\usepackage{longtable}

\usepackage{url}
\usepackage{epsfig}
\usepackage{amsmath}
\usepackage{natbib}
\usepackage{graphicx}
\usepackage{aas_macros}
\begin{document}

\title{A multiwavelength analysis of the clumpy FIR--bright sources in M33}

\author[Natale et al.]{\parbox{\textwidth}{G. Natale,$^{1,2}$$\dagger$ K. Foyle,$^{3}$$\dagger$  C. D. Wilson,$^{3}$ N. Kuno$^{4}$}\vspace{0.4cm}\\\\
\parbox{\textwidth}{
$^{1}$Jeremiah Horrocks Institute, University of Central Lancashire, Preston, PR1 2HE, UK\\
$^{2}$Max Planck Institut für Kernphysik, Saupfercheckweg 1, D-69117 Heidelberg, Germany
$^{3}$Dept. of Physics \& Astronomy, McMaster University, Hamilton, Ontario, L8S 4M1, Canada\\
$^{4}$National Astronomical Observatory of Japan, Nagano, 384 1305, Japan\\
}}
\maketitle
\label{firstpage}

\begin{abstract}

We present a multiwavelength study of a sample of far-infrared (FIR) sources detected on the {\it Herschel} broad--band maps of the nearby 
galaxy M33. We perform source photometry on the FIR maps as well as mid-infrared (MIR), H$\alpha$, far-ultraviolet and 
integrated HI and CO line emission maps. 
By fitting MIR/FIR dust emission spectra, the source dust masses, temperatures and luminosities are inferred. The sources are classified 
based on their H$\alpha$ morphology (substructured versus not-substructured) and on whether they have a significant CO detection 
($S/N>$3$\sigma$). We find that the sources have dust masses in the range 10$^2$-10$^4$~M$_\odot$ and that they present significant differences in 
their inferred dust/star formation/gas parameters depending on their H$\alpha$ morphology and CO detection classification. 
The results suggests differences in the evolutionary 
states or in the number of embedded HII regions between the subsamples. 
The source background--subtracted dust emission seems to be predominantly powered by local star formation, as indicated by a 
strong correlation 
between the dust luminosity and the dust-corrected H$\alpha$ luminosity and the fact that the extrapolated young stellar luminosity is 
high enough to account for the observed dust emission. 
Finally, we do not find a strong correlation between the dust-corrected H$\alpha$ luminosity and the 
dust mass of the sources, consistent with previous results on the breakdown of simple scaling relations at sub-kpc scales. 
However, the scatter in the relation is significantly reduced by correcting the H$\alpha$ luminosity for the age of the young stellar 
populations in the star--forming regions.

\end{abstract}
\section{Introduction}

From the morphological point of view, the dust emission observed within nearby galaxies can be roughly described in terms of 
a clumpy and a smoother diffuse component. 
\let\thefootnote\relax\footnote{$\dagger$ The first two lead authors have been co-equal contributors to the majority of the work presented in this paper.} 
At least partially, the clumpiness of the dust emission morphology is determined by 
the gas mass distribution: the clumpy component can be associated with massive giant molecular clouds (GMCs), with sizes of the order of 
$\approx 10-100$\,pc, and the diffuse component with a more diffuse interstellar medium (ISM) gas, distributed on kpc scales. 
However, peaks of emission in 
the far-infrared (FIR) maps of galaxies can also be due to the presence of strong radiation sources heating the dust locally. In addition, for a given 
dust composition and size distribution, the diffuse dust emission spectrum will depend 
on the intensity and colour of the 
kpc scale radiation field produced by all the stellar populations in a galaxy. Therefore, the
morphology of dust emission does not necessarily follow the ISM gas morphology. Understanding the intrinsic nature of the clumpy 
and diffuse components of dust emission is important since their spectral energy distributions carry 
information about both the ISM gas associated with them and the nature of the radiation sources powering them. 

Multiwavelength studies of dust emission in nearby galaxies using the {\it Herschel Space Observatory} have mainly relied on pixel-by-pixel analyses 
(e.g. Smith et al.\ 2010, Galametz et al.\ 2012, Mentuch-Cooper et al.\ 2012, Foyle et al.\ 2012, Bendo et al.\ 2012).
This method consists of convolving all the maps considered in the analyses to the same resolution, corresponding to the resolution of 
the map characterized by the largest point spread function (PSF) beam, and then comparing the emission at each wavelength or the measured dust/stellar population 
parameters at each pixel position. This method is 
advantageous for its simplicity but it has the drawback of considering simultaneously both the clumpy and the diffuse components of the 
dust emission. By including both components, pixel-by-pixel analyses are not in the position of distinguishing their properties.  

A procedure to extract and measure the dust parameters of bright FIR compact sources within nearby galaxies has been introduced
by Foyle et al. (2013, hereafter FN13) and applied to a set of multiwavelength data of the spiral galaxy M83. In order to perform the 
compact source photometry at the different wavelengths, FN13 used \textsc{getsources}, a multiwavelength source detection 
and photometry algorithm (Men'shchikov et al.\ 2012). This algorithm is designed to detect 
sources and extract the background--subtracted source photometry on {\it Herschel} maps in their original resolution, without degrading the maps.  Thus, ideally, 
the highest resolution information is preserved. Furthermore, FN13 performed a two-component dust emission SED fitting to infer the dust 
parameters and measured star formation rates (SFR) from the dust--corrected H$\alpha$ source luminosity. This procedure provided
insights on the nature of the bright FIR sources in M83 and their associated giant molecular associations (GMAs) and  showed that the 
properties of the sources do not show radial variations, which is typically found in pixel-by-pixel studies that average over both clumpy and diffuse emission simultaneously (e.g. Foyle et al.\ 2012).

In this paper we apply the same procedure developed by FN13 to a set of multiwavelength maps of the nearby galaxy M33. 
M33 is a flocculent spiral galaxy at a distance of  only 0.859 Mpc (Barker \& Sarajedini 2008). Thus, it is an ideal target to study the 
characteristics of FIR compact sources on scales even smaller than those that can be observed in M83. The dust emission in M33 has already been 
the subject of several works (e.g. Tabatabaei et al. 2014, Kramer et al. 2013, Rela{\~n}o et al. 2013, Xilouris et al. 2012, Boquien et al. 2011, 
Braine et al. 2010, Verley et al. 2010, Kramer et al. 2010).  In Rela{\~n}o et al. (2013; hereafter R13) they studied the dust properties of HII regions in M33 that they detected in H$\alpha$ emission.  The sources were classified based on their H$\alpha$ morphology and, for each source,  they measured the FIR colours and dust parameters. 

We take an alternative approach to R13, by detecting sources in the FIR emission of M33 instead of H$\alpha$.  At first, it might seem that these two approaches might be comparable. 
However, although dust in and around HII regions 
can be seen as a bright MIR/FIR compact source, in principle clumpy dust emission can arise also from sources not clearly detected in 
H$\alpha$, either because the local star formation is not intense enough for H$\alpha$ emission to be detected or because H$\alpha$ is attenuated by dust.  Furthermore, a single FIR source might be related to an entire ensemble of HII regions as well as single ones.   

In addition to the FIR emission, we also augment our data set  with mid-infrared (MIR) and UV broad--band maps as well as H$\alpha$, HI 21cm and CO line emission maps. 
MIR data can be used together with the FIR measurements on the {\it Herschel} maps 
in order to obtain a complete MIR to FIR dust emission SED of the detected sources. Dust corrected H$\alpha$ and UV luminosities can be 
used to investigate star formation on different time scales, H$\alpha$ tracing more recent star formation ($10^7$ yr)
than UV ($10^8$ yr). Finally, HI and CO gas maps can be used to elucidate the relation between the gas and the dust parameters. 

Specifically, in this work we address the following questions: \\    
1) What is the intrinsic nature of the bright FIR sources in M33? Are they associated with a single or with multiple HII regions? 
Are they mainly associated with the molecular gas detected in CO?  \\
2) How do the dust parameters vary depending on the radial distance, the H$\alpha$ morphology, the detection of molecular gas ? \\
3) Is the source dust emission heated predominantly from local young stellar populations? Is there evidence of external heating? \\
4) Assuming that the source dust mass traces the gas mass of the cloud associated with the cospatial HII regions, 
is there evidence for a scaling relation between dust mass and the recent star formation within the clouds? \\

The paper is structured as following. Section 2 presents the data set we used. In Section 3 we describe the source extraction 
technique and the photometry performed at all wavelengths. Section 4 describes the dust emission SED fitting technique applied to 
the MIR/FIR data. In Section 5 we describe the measurements of the star formation parameters. In Section 6 we explain the criteria 
adopted to select our best source sample and define subsamples of 
sources based on their H$\alpha$ morphology and the detection of CO emission. Section 7 presents the results and in Section 8 
we discuss our findings. We conclude with a brief summary.

\section{Observations}
\begin{figure*}
\centering
\includegraphics[trim=0mm 0mm 0mm 0mm,width=180mm,angle=0.]{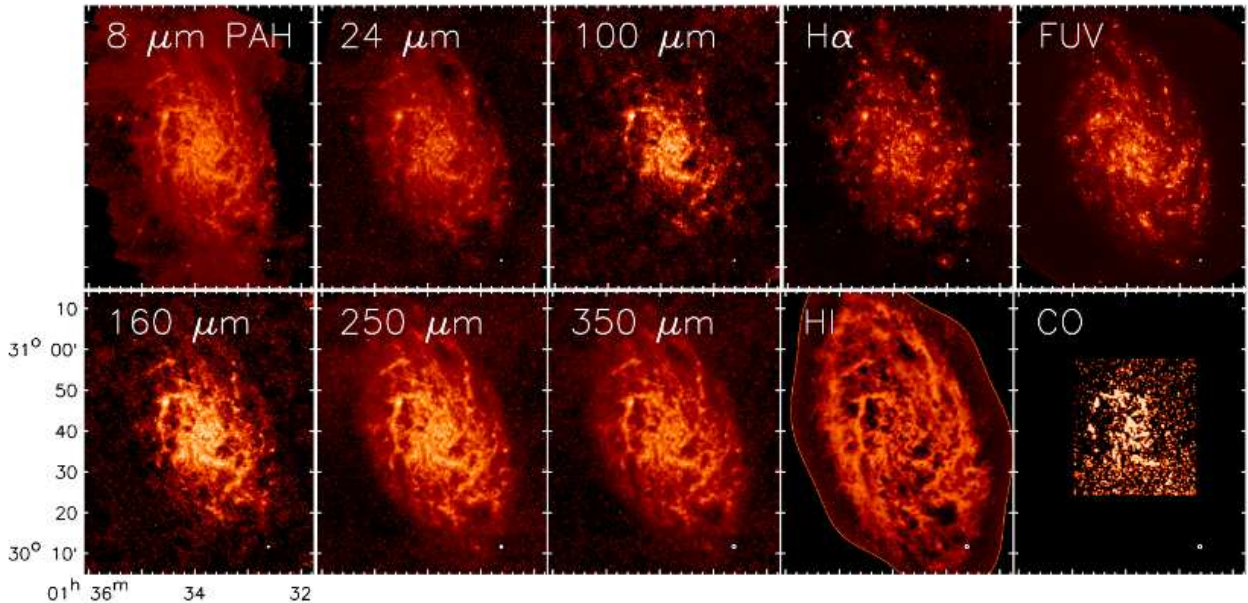}
\caption{Multi-wavelength data used in this work. Upper row: {\it Spitzer} IRAC 8$\mu$m (stellar emission subtracted) and MIPS 24$\mu$m maps,  
{\it Herschel} PACS 100$\mu$m map, H$\alpha$ map from the KPNO Schmidt telescope (Greenawalt 1998), GALEX FUV map. Lower row: {\it Herschel}
PACS 160 and SPIRE 250, 350
$\mu$m maps, VLA 21 cm line HI map, $^{12}$CO (J=1-0) integrated intensity map from the Nobeyama Radio Observatory. The diameter of the 
white circles at the right-bottom of each panel is equal to the PSF FWHM. Note that the resolution of the 8$\mu$m, 24$\mu$m, H$\alpha$,
FUV, HI and CO maps have been degraded as described in the text.}
\label{allmaps}
\end{figure*}

As in FN13, we use FIR maps to detect and extract the photometry of compact FIR bright sources but 
also perform the photometry at the same source locations on MIR and H$\alpha$ emission maps.  
In addition, we include a far-ultraviolet (FUV) map and gas maps (CO and HI) in order to trace young stellar populations bright in the 
UV and the cold gas possibly associated with the FIR sources. Here we describe the data reduction prior to the source photometry. 
All the data used are shown in Fig.\ref{allmaps}.

\subsection{FIR maps}
We use FIR maps obtained by the {\it {\it Herschel} Space Observatory}.  The images are part of the {\it Herschel} Open Time Key 
Project HerM33es ({\it Herschel} M33 Extended Survey; P.I.: C. Kramer).
We use 100 and 160~$\mu$m maps taken with the Photodetector Array Camera (PACS; Poglitsch et al.\ 2010) and 250 and 350 maps 
taken with the Spectral Photometric Imaging REceiver (SPIRE; Griffin et al.\ 2010).  The PACS and SPIRE data reduction are described 
in Boquien et al.\ (2011) and Xilouris et al.\ (2012) respectively.
The images are kept in their native resolution with a full width at half-maximum (FWHM) of the PSF of 6.8$''$, 11.4$''$, 17.6$''$,
and 24.0$''$  for the 100, 160, 250, and 350$\mu$m maps, respectively.  Note that, due to the low resolution and thus high uncertainties in the source detection method, we do not include the 500$\mu$m SPIRE map in our analysis.

\subsection{MIR maps}
In addition to the FIR maps, we also use 8 and 24~$\mu$m maps from IRAC and MIPS instruments taken from the {\it Spitzer}  Local 
Volume Legacy Survey (Dale et al.\ 2009). These MIR maps trace the warm dust and polycyclic aromatic hydrocarbon (PAH) emission.  
In the case of the 8~$\mu$m map, we subtract the stellar component of the emission, in order to extract only the dust/PAH emitting 
component. We do this by using a scaling of the IRAC 3.6~$\mu$m map, according to the relation provided by Helou et al.\ (2004): 
$F_{\nu}$(8 $\mu$m, dust) = $F_{\nu}$(8 $\mu$m)-0.232$F_{\nu}$(3.6 $\mu$m) as was done in FN13.
We degrade the resolution of the MIR maps to 6.8$''$ in order to match the resolution of the 100~$\mu$m map using the kernels 
of Aniano et al.\ (2011).   

\subsection{FUV map}
We use a FUV map from GALEX (Martin et al.\ 2005) and distributed by Gil de Paz et al.\ (2007) in order to trace UV continuum emission.  
Details on the image processing are found in Thilker et al.\ (2005).  We correct Galactic extinction using E(B-V)=0.042 from the 
NASA/IPAC Extragalactic Database that is based on the dust extinction maps from Schlegel et al.\ (1998).  The map is degraded from 
its 4.4$''$ resolution to a resolution of  6.8$''$ matching that of the 100~$\mu$m map using the kernels of Aniano et al.\ (2011).

\subsection{H$\alpha$ map}
We  trace ionized gas using an H$\alpha$ emission map from Greenawalt (1998).  The image and processing is described in 
Hoopes \& Walterbos (2000). 
As was done for the FUV map, we correct the H$\alpha$ map for Galactic extinction.  We also correct for [NII] contamination using 
[NII]/H$\alpha$ =0.05. We degrade the H$\alpha$ map from its 2$''$ resolution to a resolution of
6.8$''$, matching that of the 100~$\mu$m map, by using a Gaussian kernel.

\subsection{Molecular gas map}
In order to trace the molecular gas, we use a $^{12}$CO (J=1-0) map of M33 with 19.3$''$ resolution from the Nobeyama Radio 
Observatory (Tosaki et al.\ 2011).  We use the moment 0 integrated intensity map, which is converted to a molecular gas mass 
assuming a CO-to-H$_{2}$ conversion factor of  2 $\times$ 10$^{20}$ cm$^{-2}$(K km s$^{-1}$)$^{-1}$ (Strong et al.\ 1988).  
We multiply by a factor of 1.36 to account for helium.  We match the resolution of the map to the 350 $\mu$m map (24.0$''$) using 
a Gaussian kernel.

\subsection{Atomic gas map}
We trace the atomic gas component using HI 21 cm VLA observations (Gratier et al.\ 2010).  The integrated intensity maps are 
converted to a gas mass assuming a conversion factor of 1.8$\times$10$^{18}$ cm$^{2}$/(K km s$^{-1}$) (Rohlfs \& Wilson 1996).  
For consistency with the CO map, the atomic gas map is convolved from its native resolution of 17$''$ to the resolution of the 
350$\mu$m map. 

\section{Source Detection and Photometry}
In this section we describe the source detection and photometry method applied to the FIR data set (\S3.1) as well as the photometry 
performed on all the other maps at the location of the detected FIR sources (\S3.2). 

\subsection{FIR source detection and photometry}
We detect and extract the photometry of compact sources in the FIR maps using \textsc{getsources}, a multiwavelength and multiscale 
source detection tool (Men'shchikov et al.\ 2012). We use this tool because it can account for the different angular resolutions of the 
FIR maps, perform background subtractions in the vicinity of the sources and account for source blending.  A description of the code 
and its application to the nearby galaxy M83 is provided in FN13.  Here we briefly describe the main steps performed by this code.
 
Rather than detecting the sources in the original images, \textsc{getsources} creates single-scale detection 
maps by convolving the images with a series of gaussians and then subtracting the convolved images one from the other (see formula 1
of Men'shchikov et al.\ 2012). 
For a given spatial scale, the maps at each wavelength 
are combined to produce 
a single scale detection image.  A source can then be detected and tracked through the different scales. The scale where the 
source is the brightest provides an initial estimate of the source size and shape.  Once the sources have been detected, the original 
observations are used to extract the photometry of the source.  A background in the vicinity of the source is also interpolated and 
subtracted at each wavelength.
 
Before the source extraction process, we first align the images to the same grid and convert them to MJy sr$^{-1}$, with a pixel size 
of 1.4$''$. Observational masks are made, which mark the extent of the 
image over which \textsc{getsources} searches for sources. In FN13, the MIR 24$\mu$m and H$\alpha$ maps were also included in the 
\textsc{getsources} source extraction process.  However, as described below, we perform aperture photometry on the ancillary bands 
because, at the spatial scales observed in M33 (i.e. 28 pc at 100$\mu$m
), the emission morphology at those wavelengths may differ considerably from that observed in the FIR.
 
After the extraction process, \textsc{getsources} returns a table with the source properties including the source size, position, 
total flux and peak flux at each wavelength. \textsc{getsources} also returns a monochromatic detection significance, which is 
determined using 
the ratio of the peak flux of the source to the standard deviation in an annulus surrounding the source on the detection maps.  

Since many sources are only marginally detected, we set a minimum detection significance threshold to select a sample of sources to be included 
in the analysis. Specifically, we required a detection significance higher than 3$\sigma$ for the 100$\mu$m and 160$\mu$m bands, which 
are those with the highest angular resolution in the FIR data set.  
 
\subsection{Aperture photometry at other wavelengths}
After measuring the FIR fluxes for the detected sources on the {\it Herschel} maps, we perform aperture photometry at the same positions on 
the other maps of our data set to obtain the corresponding total fluxes. We used aperture photometry, rather than using 
\textsc{getsources} directly on the entire data set, because the morphology observed for the young stellar populations, ionized and cold gas, 
dust and PAH emission in the area covered by the FIR sources can be intrinsically different from the FIR morphology.
For example, several close HII regions bright in H$\alpha$ might be seen as a single FIR source. 
The aperture sizes correspond to the source footprints inferred by \textsc{getsources} at 100$\mu$m for the H$\alpha$, FUV, 8 and 
24$\mu$m and at 350$\mu$m for the HI and CO maps. The choice of these apertures is consistent with the angular resolutions to which 
we convolved the different maps. 

In this way, we aim to measure the amount of H$\alpha$, UV, 8 and 24$\mu$m that presumably is physically associated with detected FIR 
sources. In the case of HI and CO emission, we did not seek to measure the fluxes directly associated with the FIR compact sources, 
since these maps present a rather smooth emission at those locations, with no or only marginal evidence of compact sources coincident
with the detected FIR sources. This is partially due to the low resolution of the gas maps and partially to 
the fact that a large fraction of the gas could be in the form of a background component not associated with the FIR bright source. 
Since it was not possible to estimate a GMC mass for the FIR sources, we used the HI and CO maps only to measure gas mass 
column density in the areas where the sources are detected. This allows us to investigate if there is any relation between the source 
dust parameters and the local gas column density. 

While performing the aperture photometry on the H$\alpha$, FUV, 8 and 24$\mu$m maps, we also subtracted a background emission component, 
which we estimated in circular apertures of different size selected by-eye throughout the galaxy disc. We also estimated the
variation of the brightness in the background apertures, which was used to estimate the photometric uncertainties 
for each measurement. We did not apply a background subtraction for the measurements on the gas maps since our goal was 
to measure the total gas surface density in the region of the sources.  
By including a background subtraction for the FIR emission (see the previous section) and for the MIR and star formation tracers, 
we assume that we are removing a diffuse emission component that is not directly associated with the FIR compact source. Given the large uncertainties with any approach, we chose to apply the simple approach outlined here.   

\section{Dust/PAH emission SED Fitting}

We fit the inferred MIR and FIR fluxes with the dust/PAH emission SED fitting procedure 
developed by Natale et al. (2010). As this procedure was already used in FN13, here we only summarize the method and 
we refer to Natale et al. (2010) and to FN13 for further details. \\
The observed fluxes are fit by two dust emission components: a photo-dissociation region (PDR) emission component, having a fixed spectral shape
and peaking at about 70$\mu$m, and
a diffuse dust/PAH emission component, representing the emission from diffuse dust heated by interstellar radiation 
fields (ISRF) of lower intensity compared to those powering the PDR emission. In the latter case, the spectral shape of the emission is not 
fixed but the best--fitting model is found within a library of 361
dust emission templates. The SED shape of these templates varies depending on the heating radiation intensity and optical/UV colour. 
The fitting parameters are: $\chi_{\rm UV}$, the amplitude of the UV part of the radiation field heating the diffuse dust 
in units of the Mathis et al. (1983) curve (representing the local ISRF); $\chi_{\rm col}$, the optical/UV ratio of the heating radiation spectra; 
M$_{\rm SED}$ the diffuse dust mass; F$_{\rm 24}$ the fraction of $24\mu$m flux due to the PDR emission template. 
Fig.~\ref{sed_fits} shows the SED fit for four example sources.

Apart from the 
fitting parameters, the SED fitting procedure also provides the diffuse dust temperature T$_{\rm SED}$,  which is the temperature of the 
modified blackbody with an emissivity index of $\beta$=2 reproducing the FIR bump of the best--fitting diffuse dust emission template. 
For comparison, we also perform standard one--component modified blackbody fits (again with $\beta=2$) to the observed SEDs 
by including only the FIR {\it Herschel} maps (see FN13 for details). Using this fitting technique, we infer the dust mass M$_{\rm BB}$ and temperature T$_{\rm BB}$.  
Note that T$_{\rm BB}$ can sometimes be different than T$_{\rm SED}$, because in the first case
the entire flux in the FIR bands has to be reproduced by the best--fitting curve. Instead, in the two component PDR/diffuse dust SED fit, 
part of the flux in the FIR bands, especially at 100$\mu$m and 160$\mu$m, can be assigned to the warm PDR component.  
Since the relative fraction of FIR emission fit by the PDR component is higher at shorter FIR wavelengths, 
the T$_{\rm BB}$ temperatures are often warmer than that measured by T$_{\rm SED}$.

\section{Star formation parameters}

H$\alpha$ and FUV emission fluxes need to be corrected for both Galactic extinction and internal dust attenuation. Galactic
extinction is due to the Milky Way dust present in the line of sight of the observed galaxy. This is corrected using the extinction
value taken from Schlegel et al. (1998), as described 
in \S2.3 and \S2.4. Internal dust attenuation is due to both the local dust close to the star--forming regions and the dust in 
the diffuse ISM of the observed galaxies. To correct for internal attenuation, we use empirical relations based on using the observed $24\mu$m 
flux as a proxy for the amount of internal attenuation. Specifically, we use the relation provided by Calzetti et al. (2007) to correct
the H$\alpha$ luminosity: 
\begin{equation}
{\rm L(H}\alpha,{\rm cor)=L(H}\alpha)_{{\rm obs}}{\rm +0.031L(24}\mu\rm{m)}
\end{equation}
where L(H$\alpha$)$_{{\rm obs}}$ and L(H$\alpha$,cor) are the observed and intrinsic H$\alpha$ luminosities respectively, while 
L(24$\mu$m)=$\nu$$_{24\mu m}$ L$_\nu(24\mu m)$ with L$_\nu(24\mu m)$ the observed luminosity density at 24$\mu$m. Throughout the text, 
we will refer to the term 0.031L(24$\mu$m) as the obscured H$\alpha$ luminosity or L(H$\alpha$, obscured) and 0.031L(24$\mu$m)/L(H$\alpha$,cor) as the ``fraction of obscured H$\alpha$ luminosity''.

To correct the FUV, we consider the relation D10 of Leroy et al. (2008), which can be used to calculate the SFR from the 
observed FUV and 24$\mu$m luminosities. We used this relation divided by 0.68$\times$10$^{-28}$, the coefficient used to convert the 
dust corrected FUV luminosity into SFR:  

\begin{equation}
 {\rm L}_{\rm \nu,cor}{\rm(FUV)=L}_{\nu,{\rm obs}}{\rm (FUV)+3.14}\times10^{-14}{\rm L(24}\mu\rm{m)}
\end{equation}
where L$_{\nu,{\rm obs}}$(FUV) and L$_\nu$(FUV) are the observed and intrinsic FUV luminosity densities.   

We do not convert the H$\alpha$ and FUV luminosities into SFR, since at these spatial scales the assumptions of 
constant SFR and complete initial mass function (IMF) sampling are not applicable (e.g. Calzetti, Liu \& Koda 2012). 
However, we use the dust corrected H$\alpha$ and FUV luminosities to determine a dust corrected UV-to-H$\alpha$ flux ratio. In the simple 
assumption of a single--age stellar population (see also Rela{\~n}o \& Kennicutt 2009 for a similar approach), the intrinsic 
UV-to-H$\alpha$ ratio can 
be used to derive an age for the stellar population observed within each FIR source. This is because for a single--age stellar population the 
H$\alpha$ luminosity fades away within 10$^7$\,yr while the FUV takes of the order of 10$^8$\,yr to disappear. The overall result is that 
the UV-to-H$\alpha$ ratio shows a monotonic increase as a function of the stellar population age. 

To convert UV-to-H$\alpha$ ratio into
stellar population ages, we considered the standard output of the software \textsc{Starburst99} (Leitherer et al. (1999) for an instantaneous burst of star formation, 
which provides the emission spectra of a single--age stellar population as a function of age. In the calculation we assumed a 
Kroupa (2001) IMF, with a slope of -1.3 between 0.1 and 0.5 M$_\odot$ and -2.3 between 0.5 and 
100 M$_\odot$,
and solar metallicity $Z$=0.02.  
After extracting the UV/H$\alpha$ ratio as a function of age from the \textsc{Starburst99} output, we use this theoretical relation
to convert the observed UV/H$\alpha$ ratio of each source into an age. 
Then, we use this information to correct the source H$\alpha$ luminosity for age effects. That is, we 
derive the H$\alpha$ luminosity that each source would have had if they all were $t$=6$\times 10^6$\, yr old and call this age-corrected H$\alpha$ luminosity L(H$\alpha$,age). To this purpose, the 
correction one needs to multiply by the observed dust corrected H$\alpha$ luminosity is equal to L(H$\alpha$,$t_o$)/L(H$\alpha$,$t_s$), 
where L(H$\alpha$,$t$) are the H$\alpha$ luminosities calculated by the \textsc{Starburst99} run at time $t_o$=6$\times$10$^6$\,yr and 
at the source age $t_s$. As discussed in Section \S7.4, the age corrected H$\alpha$ luminosity allows a fairer comparison 
between the sources when discussing the star formation associated with them.  

The above procedure which is used to determine an age-corrected H$\alpha$ luminosity is not free of caveats. First, 
while the H$\alpha$ correction using the Calzetti et al. (2007) formula has been calibrated also on spatial scales similar to those 
analysed in this work, the FUV dust correction formula from Leroy et al. (2008) has been calibrated on slightly larger scales of the order 
of $\approx 1$kpc. 
Also, the assumption of 
single--age stellar population is not completely correct. For example, in the cases where more H$\alpha$ sources are seen
within a single FIR aperture (see the next section), it is probable that the observed stellar populations are not born exactly during the 
same star formation event. A more accurate analysis could be done by using multiwavelength data of the 
stellar emission for the young star clusters in M33 in order to derive the star formation history from spectral SED fitting 
(see Sharma et al. 2011). However, this kind of analysis is beyond the scope of this paper.   

\section{Source Selection and Classification}
\textsc{getsources} detects 600 sources in the FIR maps.  However, for our analysis we select only the sources well-detected in the 
FIR bands with highest angular resolution. That is, we consider only the detected sources having a monochromatic 
detection significance at 100$\mu$m and 160$\mu$m higher than 3$\sigma$. This reduces the sample to 348 sources. 
We also make another cut to the source list by selecting only 
those having a minimum $\chi^{2}$ in the dust emission SED fitting lower than 10 (228 sources left). Finally, we consider only the 
sources in the field of view common to all the maps. This leaves us with 183 well-detected sources with reasonably good 
SED fits and with photometry at all wavelengths. 

It is important to note that the sources rejected due to our detection significance criteria and $\chi^{2}$ cut are not located preferentially in particular regions within the galaxy. 
There are number of reasons which may cause a source to have a low detection significance: 1) the source may be intrinsically faint and/or in a region with high background noise, 
2) the source may be in a crowded region where 
deblending is more uncertain and 3) spurious sources may be detected by \textsc{getsources} due to large emission enhancements. On the other hand the sources which we discarded because of the low 
quality of the SED fit ($\chi^{2}$ cut) typically have one or two data points which make the entire SED quite irregular. This probably reflects difficulties in the 
extraction/photometry for those sources, especially for crowded regions and at longer wavelengths where deblending is more problematic.

\begin{figure*}
\centering
\includegraphics[trim=0mm 0mm 0mm 0mm,width=130mm,angle=0]{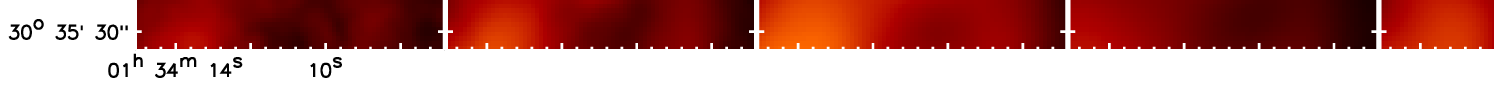}
\includegraphics[trim=0mm 0mm 0mm 0mm,width=130mm,angle=0]{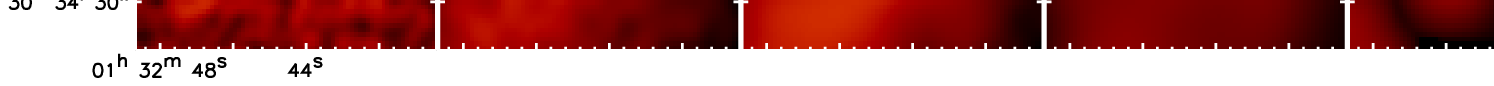}
\includegraphics[trim=0mm 0mm 0mm 0mm,width=130mm,angle=0]{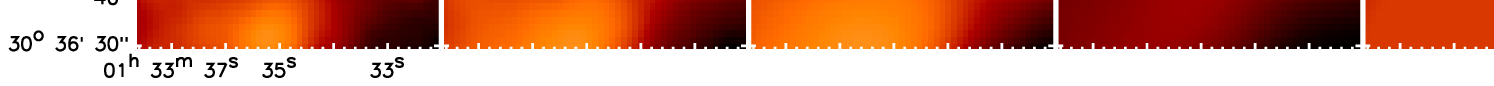}
\includegraphics[trim=0mm 0mm 0mm 0mm,width=130mm,angle=0]{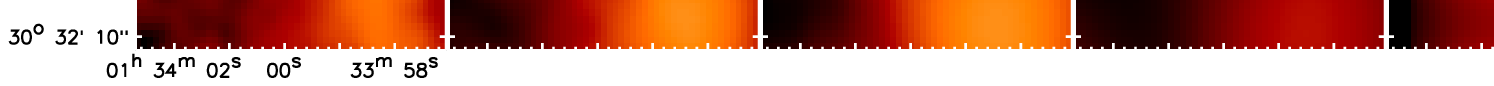}
\caption{Example of NOSUB (top), SUB (second from top), HighCO (second from bottom) and LowCO (bottom) sources detected on the FIR {\it Herschel} maps. 
The left--hand panels show the emission at each wavelength, 
delimited by the elliptical green contours (semi-axis equal to FWHM), and 
the right--hand plot show the observed source dust emission SED fit by using the PDR (dotted) and diffuse dust (dashed) emission components.}
\label{sed_fits}
\end{figure*}

\begin{figure*}
\centering
\includegraphics[trim=0mm 20mm 0mm 0mm,width=50mm,angle=-90]{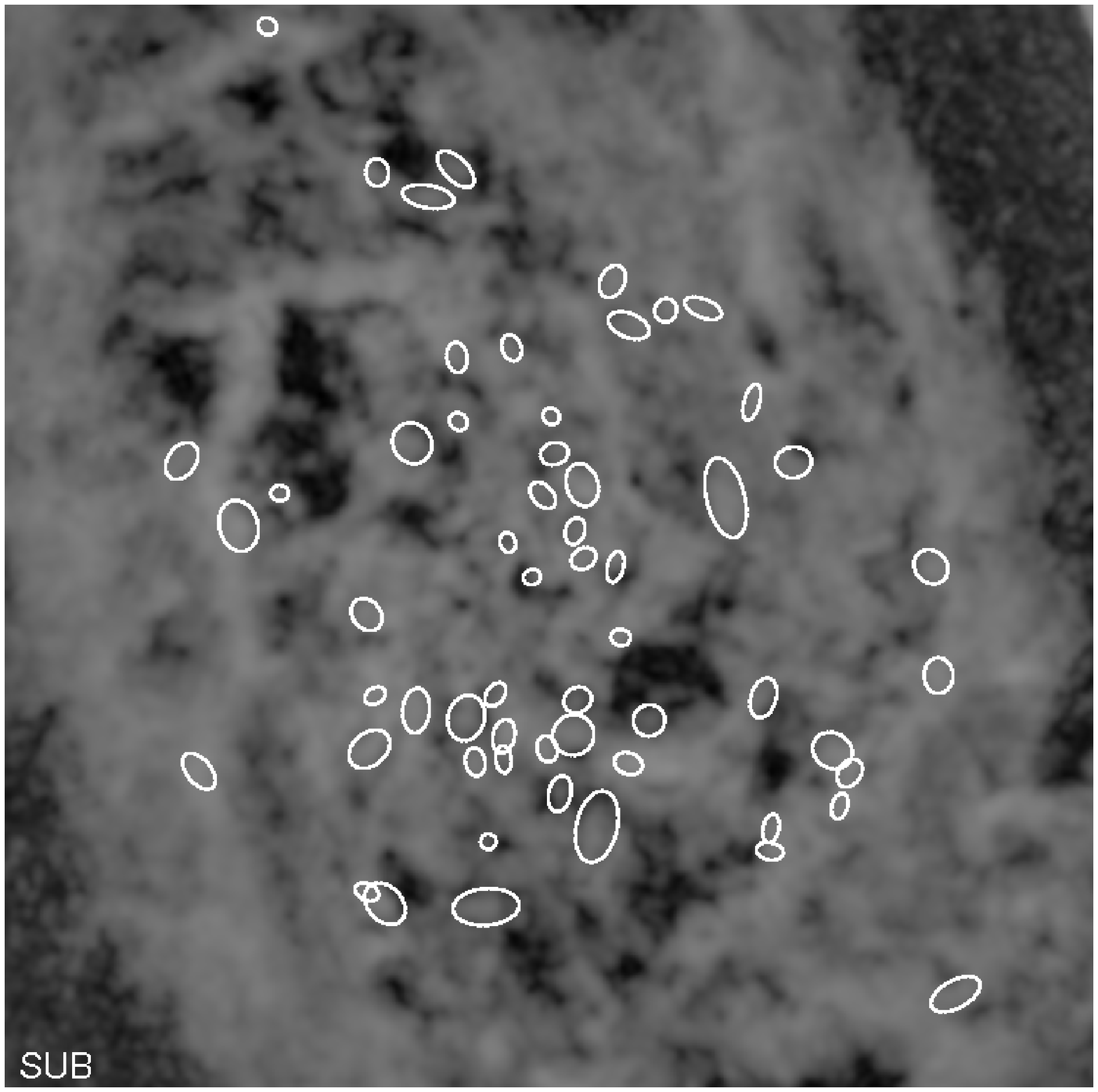}
\includegraphics[trim=0mm 20mm 0mm 0mm,width=50mm,angle=-90]{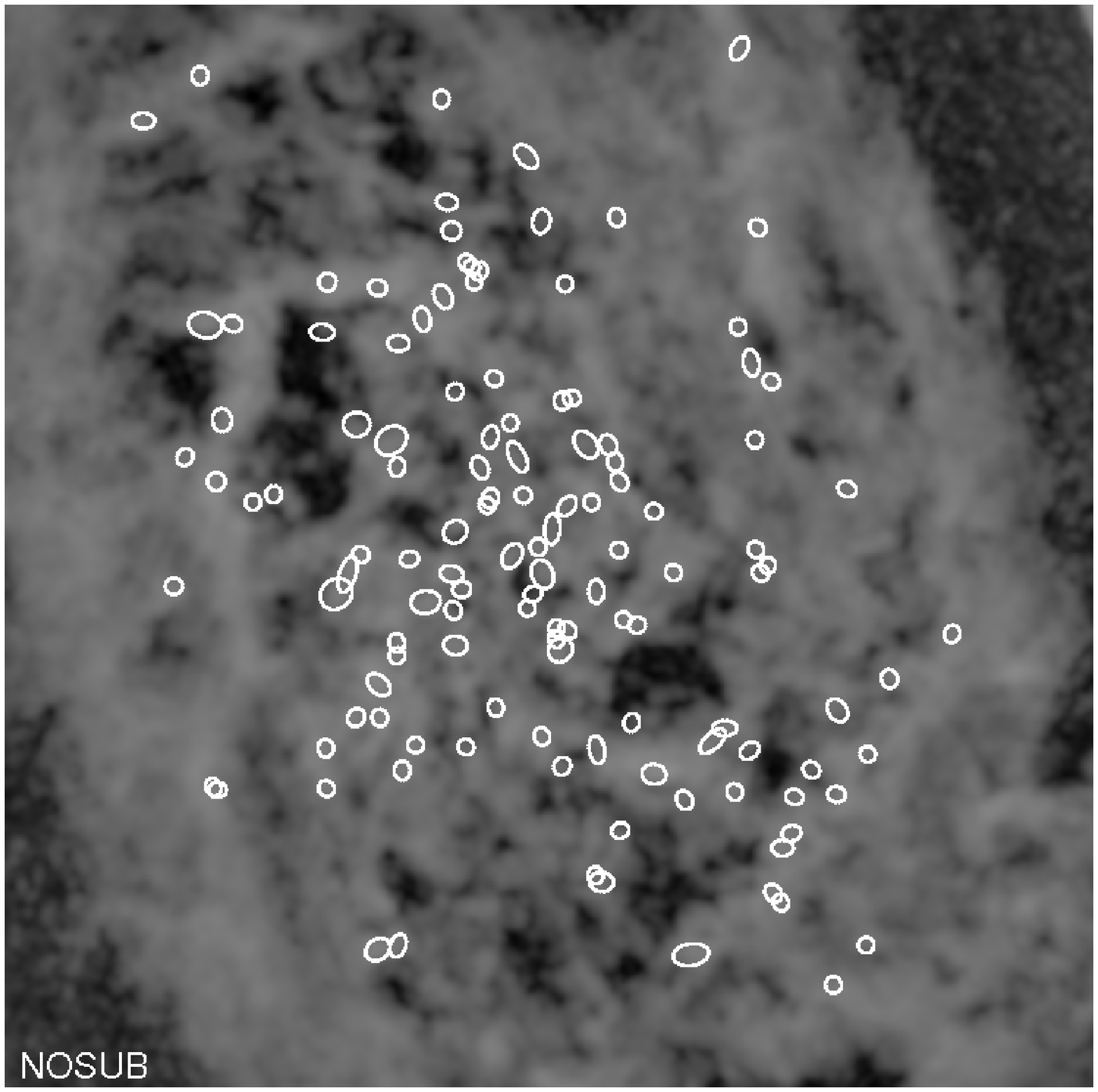}
\includegraphics[trim=0mm 20mm 0mm 0mm,width=50mm,angle=-90]{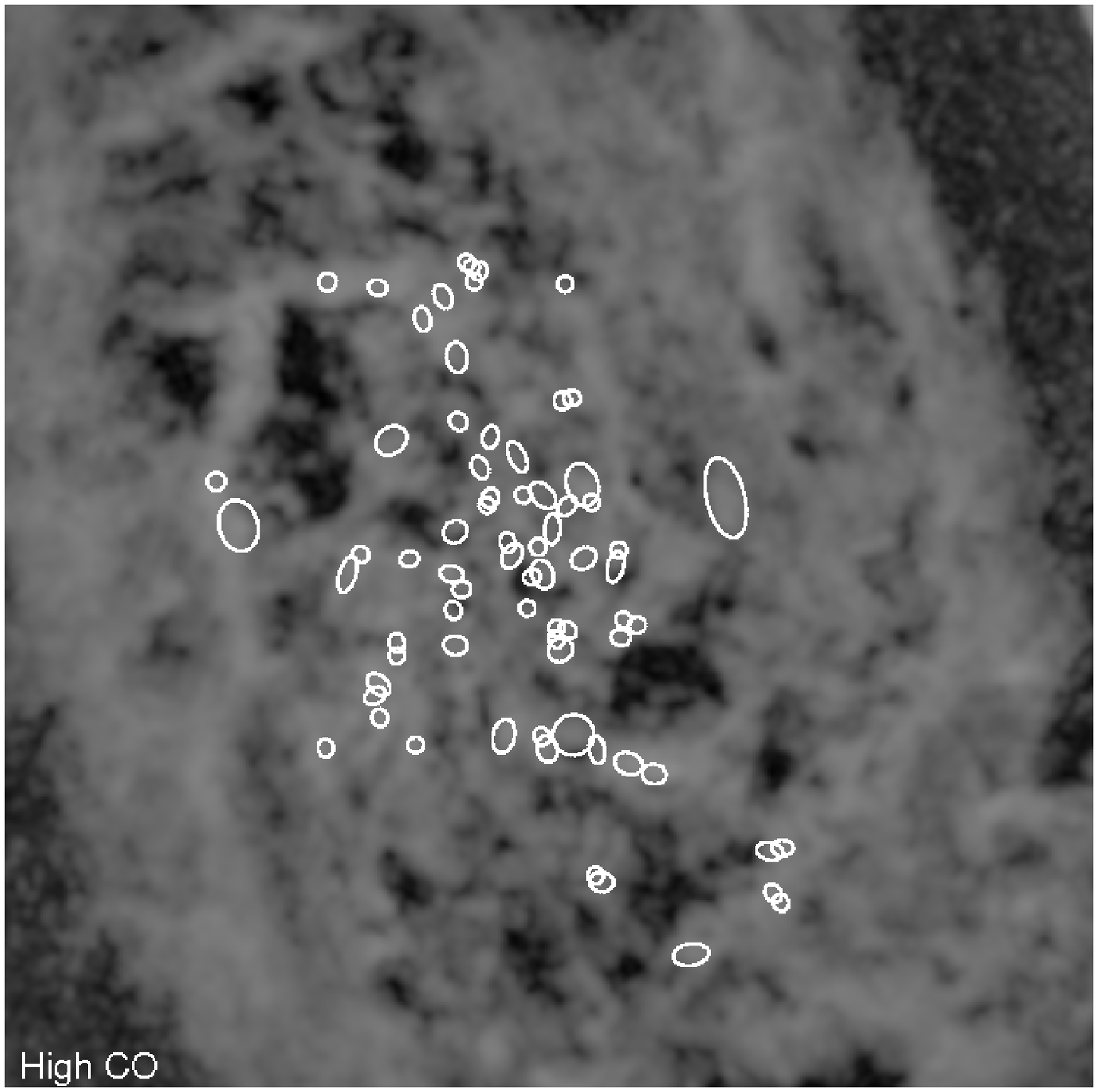}
\includegraphics[trim=0mm 20mm 0mm 0mm,width=50mm,angle=-90]{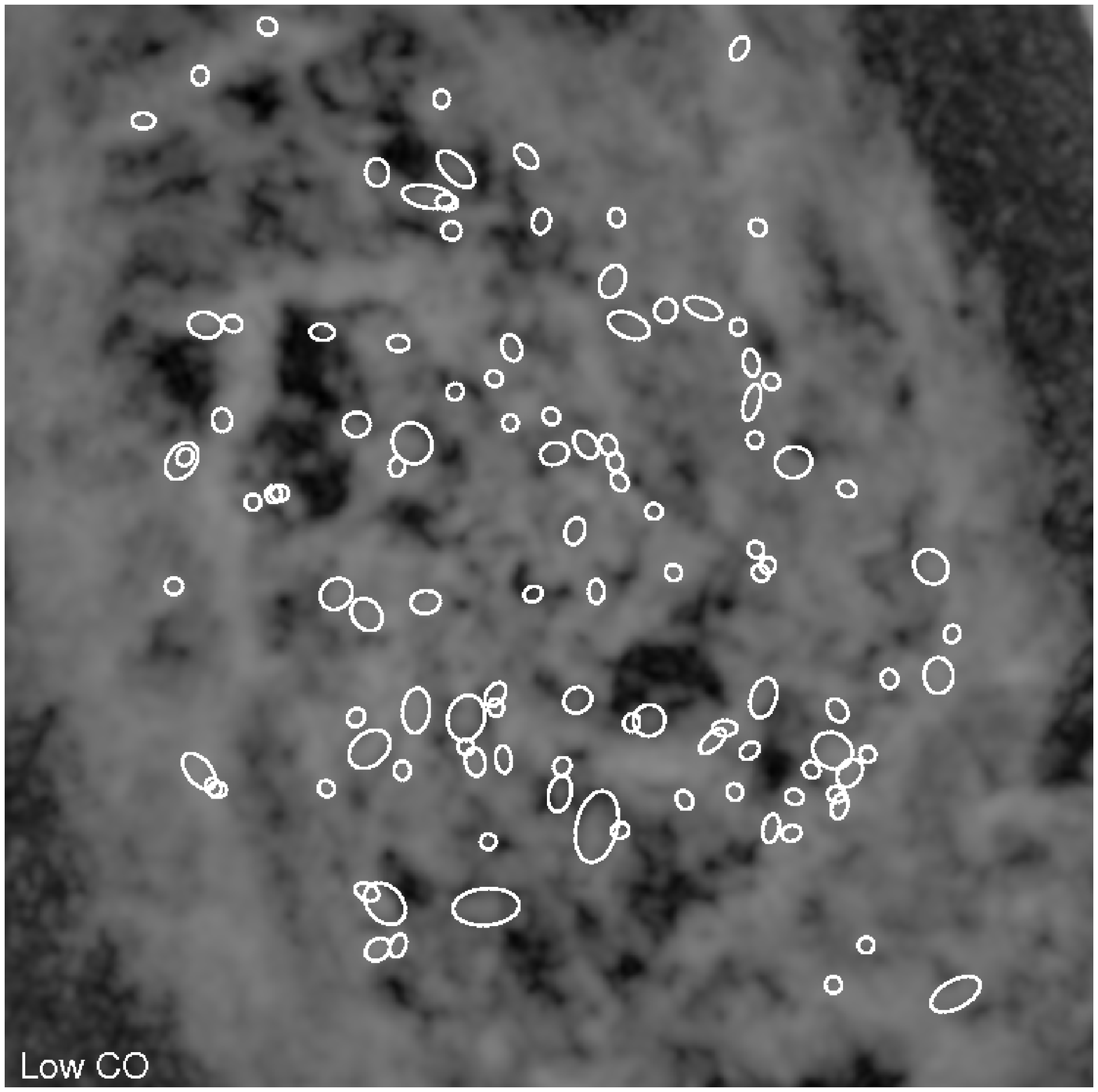}

\caption{ Source footprints at 160$\mu$m overplotted on the inner regions of the HI map with CO coverage (see bottom right--hand panel of Fig~\ref{allmaps} for extent). }
\label{SUB_CO_sources}
\end{figure*}

In order to elucidate the effects on the dust parameters related to the characteristics of the young stellar populations and the cold 
gas, observed at the same positions of the FIR sources, we subdivide our best source sample into sub-categories. 
After a visual inspection of the multiwavelength emission in the regions of each source, we differentiate the sources based on their
H$\alpha$ morphology and the amount of or lack of CO emission. By visually examining 
the H$\alpha$ emission within each FIR source footprint on the unconvolved H$\alpha$ map,
we classify the sources as non-substructured (NOSUB) or substructured (SUB). NOSUB sources are characterized by 
a single peak in H$\alpha$ while SUB sources show a more complex morphology, 
characterized by multiple peaks or an irregular smoother emission. In physical terms, the difference in morphology might depend on whether
the FIR source is associated with a single or with multiple HII regions.  The upper two panels of Fig.~\ref{SUB_CO_sources} show the location of the NOSUB and SUB sources. Examples of NOSUB and SUB sources are shown in the upper two rows of Fig.\ref{sed_fits}.

The visual inspection of the multiwavelength data also led us to differentiate between sources which are well-detected or not in CO, 
which we label ``HighCO'' and ``LowCO''. Specifically, a source is classified as HighCO if, within the FIR source footprint, there is a 
detection in the molecular gas map that exceeds three times the rms noise level in the map (that is, if there is at least a position 
where the molecular gas column density is higher than 3$\times$10 M$_{\odot}$pc$^{-2}$). In the opposite case, the source is classified as LowCO.   The lower two panels of Fig.~\ref{SUB_CO_sources} shows the location of HighCO and LowCO sources.  The lower two rows of  Fig.\ref{sed_fits} show examples
of a HighCO and LowCO source. 

The number of sources in each group can be found in Table \ref{table_class}. Note that each detected source has two labels, one for each 
type of  source classification. For example a given source might be simultaneously a NOSUB and HighCO source. 
Hereafter, the total source sample will be referred as ``ALL'' sample.  In Table \ref{table_prop} we provide a list of the properties 
inferred for each source based on the SED fitting, modified blackbody fitting, aperture photometry and corrected H$\alpha$ and UV 
luminosities. The measured fluxes in each band, the results of the two component SED fitting and the classification labels for the 
sources belonging to our best sample can be found in the appendix. 

\begin{table}
\begin{center}
\caption{Source classification}
\begin{tabular}{l | p{4cm} | l}
\hline\hline
Name & Description & Number\\
\hline
NOSUB & single H$\alpha$ source & 125\\
\hline
SUB & substructured in H$\alpha$ & 58\\
\hline
HighCO & well detected in the molecular gas map (S/N $>$ 3$\sigma$)& 73\\
\hline
LowCO & not or only marginally detected in the molecular gas map (S/N $<$ 3$\sigma$) & 110\\
\hline
\end{tabular}
\label{table_class}
\end{center}
\end{table}

\begin{table}
\begin{center}
\caption{Inferred source properties}
\begin{tabular}{l | p{5cm}}
\hline\hline
Label & Description \\
\hline
\hline
SED fitting parameters \\
\hline
M$_{\rm SED}$ & total dust mass of the source inferred from the two--component SED fit \\
\hline
T$_{\rm SED}$ & cold dust temperature of the source inferred from the two--component SED fit\\
\hline
T$_{\rm BB}$ & dust temperature of the source inferred from a modified blackbody fit of the FIR bands with $\beta$=2 \\
\hline
L$_{\rm TOT}$ & total dust luminosity inferred from the two-component SED fit \\
\hline
L$_{\rm PDR}$/L$_{\rm TOT}$ & fraction of total dust luminosity attributed to the PDR component \\
\hline
Star formation parameters \\
\hline
L(H$\alpha$,cor) & H$\alpha$ luminosity corrected for dust attenuation \\
\hline
L(H$\alpha$,obscured)/L(H$\alpha$,cor) & fraction of dust obscured H$\alpha$ luminosity \\
\hline
L(FUV, cor)/L(H$\alpha$,cor) & ratio of the dust--corrected UV luminosity to the dust--corrected H$\alpha$ luminosity \\
\hline
L(H$\alpha$,age) & H$\alpha$ luminosity corrected for dust attenuation and age effects \\
\hline
$\Sigma_{\rm gas}$ & local total gas (HI $+$ H$_{2}$) column density in the vicinity of the source \\
\hline
\end{tabular}
\label{table_prop}
\end{center}
\end{table}

\section{Results}
In this section we present the results we obtained from our analysis. In \S7.1 we show the distributions of the inferred dust and star 
formation parameters for the entire source sample and the different source categories. In \S7.2 we show how these parameters vary with radial distance. In \S7.3 
we show the results concerning the dust heating in the FIR sources and in \S7.4 those related to the star formation and gas 
associated with them. \\

\subsection{Inferred Properties of Complete Sample and Subsamples}
\begin{figure*}
\centering
\includegraphics[trim=0mm 0mm 0mm 0mm,width=110mm]{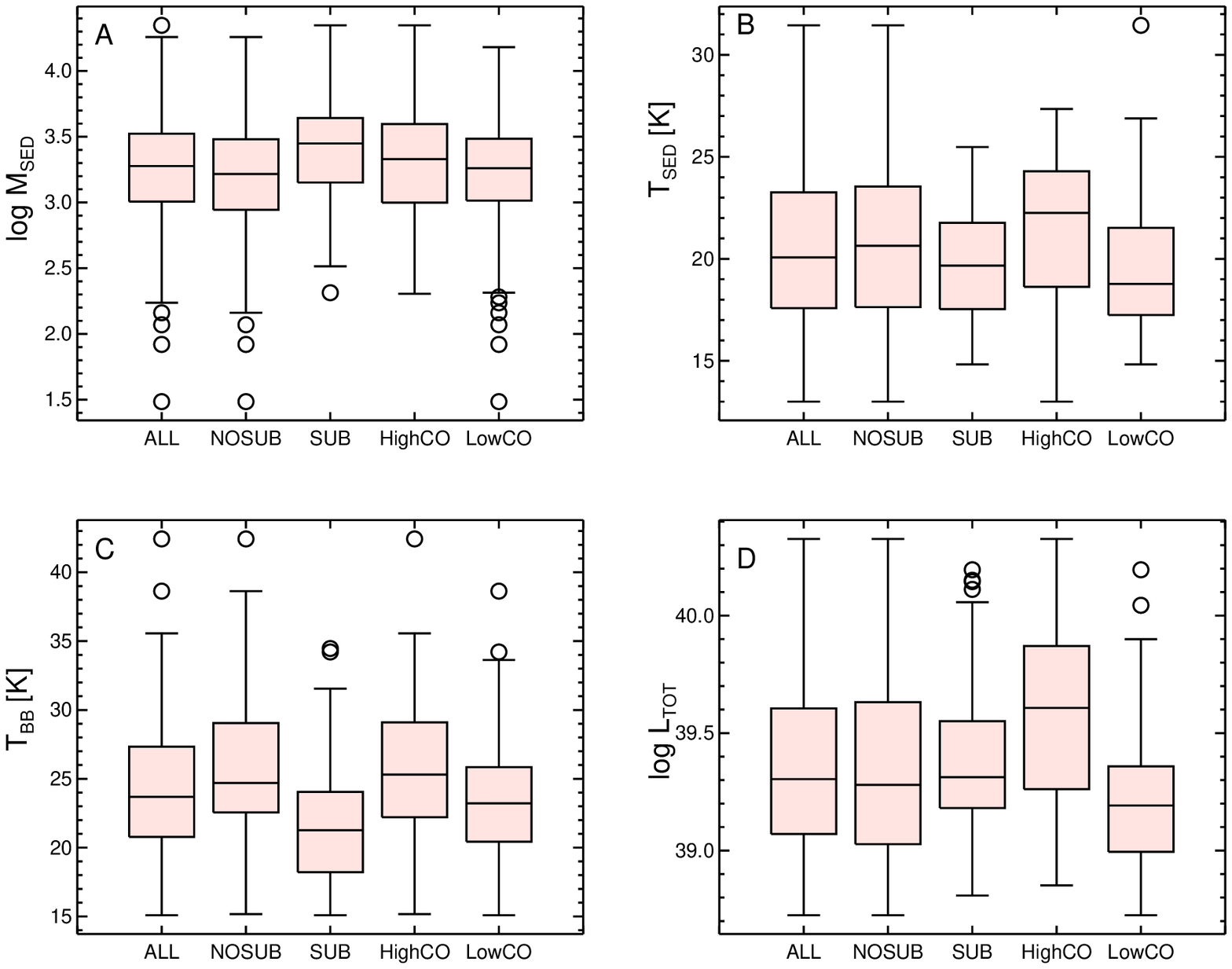}
\includegraphics[trim=0mm 0mm 0mm 0mm,width=110mm]{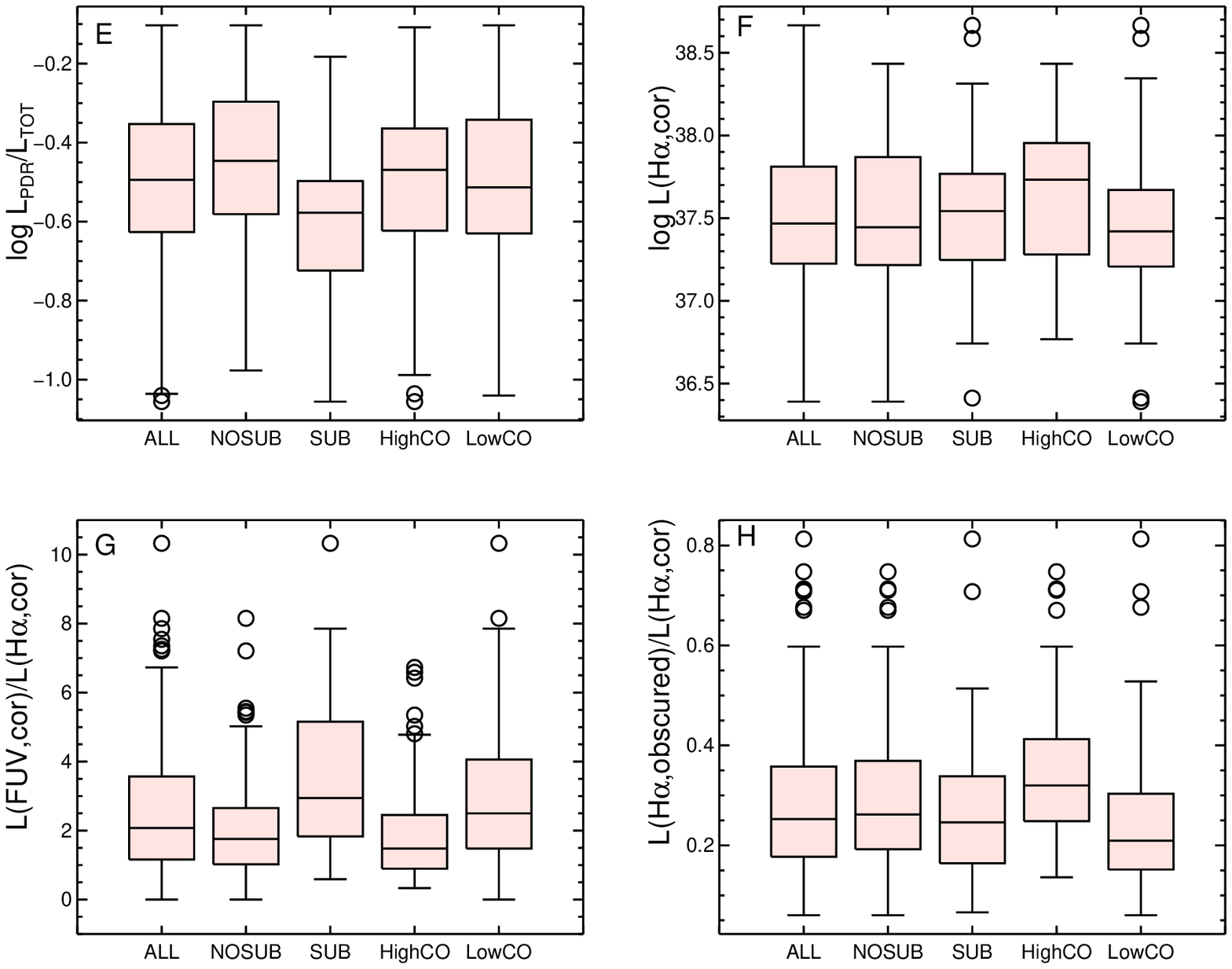}
\includegraphics[trim=0mm 70mm 0mm 0mm,width=110mm]{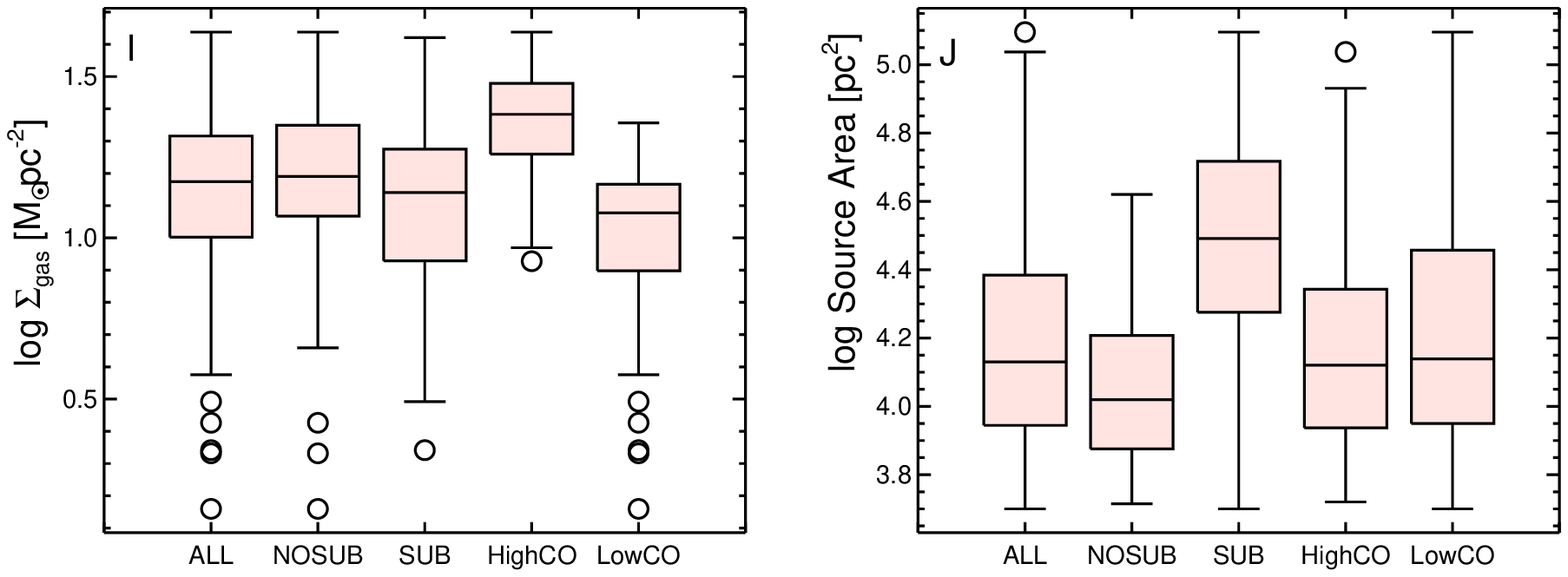}
\caption{Box-and-whisker plots for the measured source parameters. (a) Dust mass, (b) two component SED fit temperature T$_{\rm SED}$, 
(c) modified blackbody dust temperature T$_{\rm BB}$, (d) Dust luminosity, (e) PDR--to--total dust emission luminosity ratio, (f) dust 
corrected H$\alpha$ luminosity, (g) dust corrected FUV to H$\alpha$ luminosity ratio, (h) obscured to total H$\alpha$ luminosity ratio,
(i) total gas mass surface density within the 350$\mu$m source footprint, (j) deprojected source footprint area at 100$\mu$m}
\label{fig_boxplots}
\end{figure*}

In order to show the distributions of inferred parameters and compare them for the different subsamples of compact sources, we created 
box-and-whisker plots which are shown in the panels of Fig.~\ref{fig_boxplots}. The solid line in the centre of the boxes shows the median 
value and the box delineates the 25th and 75th quartiles. The lines extend to either the maximum and minimum values or to 1.5 
times the 75th and 25th quartiles. If there are values beyond the later range, they are denoted with open circles.

We first consider the general properties of the entire sample (labelled `ALL' in the box-whisker-plots).  In general, we find that the 
compact sources have dust masses between 10$^{2}$-10$^{4}$ M$_{\odot}$ and a median value of 1890 M$_{\odot}$ (panel A).  This implies gas 
masses of approximately 10$^{4}$-10$^{6}$M$_{\odot}$ assuming a constant gas-to-dust ratio of the order of 100.
Due to the proximity and superior spatial resolution of M33, the compact sources are an order of 
magnitude smaller than that detected in M83 (FN13), with median radii of 60 pc at 100$\mu$m resolution.  Thus, these compact sources likely 
represent a combination of GMCs and larger giant molecular associations (GMAs). 

In panels B and C of Fig.~\ref{fig_boxplots} we show the dust temperatures inferred from the two component SED fitting procedure (T$_{{\rm SED}}$) and 
the one--component modified blackbody fit (T$_{{\rm BB}}$).  
The temperatures both fall in the range of 15-35 K.
The T$_{{\rm SED}}$ temperatures tend to be lower (median value of 20 K) in comparison to T$_{{\rm BB}}$ temperatures (median value of 24 K). 
Differences between the two temperatures are expected because of the different way to fit the SEDs, with one or two components 
(see also \S4). 

The compact sources have dust luminosities of $\approx$10$^{39}$-10$^{40}$ erg s$^{-1}$ (panel D).  We note that there are no sources fainter 
than 10$^{38.5}$ erg s$^{-1}$ in our sample. However, some of the fainter sources detected by \textsc{getsources} are excluded because of the assumed 
lower limit for the S/N ratio (see \S6).  In general, the diffuse component of the SED fit dominates the total luminosity and we find a 
median value of the PDR-to-total dust luminosity ratio of 0.32 (panel E).

The compact regions have dust-corrected H$\alpha$ luminosities in the range of 10$^{36.5}$-10$^{38.5}$ erg s$^{-1}$ (panel F).  
The sources have a median FUV-to-H$\alpha$ 
luminosity ratio of 2.1 (panel G) and the fraction of obscured H$\alpha$ luminosity has a median value of 0.25 (panel H). 
The median total gas column density within the source areas is 15 M$_\odot$ pc$^{-2}$ (panel I).
 
Looking at Fig.~\ref{fig_boxplots}, we find that there are differences between source subsamples.  We examine to what extent the properties of the sources in the SUB-NOSUB and HighCO-LowCO subsamples have similar distributions, 
by employing a two-sided Kolmogorov--Smirnov test (hereafter KS-test).  
Following Hughes et al. (2013), we take each property and 
repeat the KS-test 500 times sampling the property within 3$\sigma$ of the uncertainties.  The median $p$-values are calculated for each 
test and listed in Table 3.  We consider two distribution statistically different when $p <$ 0.05, which is a 95\% confidence interval.  
We find that the samples do show some statistically significant differences.
We first compare the compact sources that show substructure in H$\alpha$ (SUB sample) to 
those sources that only show a single peak in H$\alpha$ (NOSUB sample). 

By performing KS-tests of the NOSUB and SUB subsamples we find that the sub-samples show different PDR-to-total dust luminosity ratio and FUV-to-H$\alpha$ luminosity ratios.  The SUB sources  have lower fractions of PDR-to-total dust luminosity with median values of 0.26 in 
comparison to 0.36 for the NOSUB sources.  The SUB sources have a median FUV-to-H$\alpha$ luminosity ratio that is 1.5 times higher than 
that of the NOSUB sources.   The 
 SUB sources also have much greater extents with areas that are close to three times greater than that of the NOSUB sources (see panel J of Fig.~\ref{fig_boxplots} ).   

The differences between NOSUB and SUB sources point to a scenario wherein the SUB sources are larger sources, containing a higher 
fraction of diffuse cold dust, as shown by their lower L$_{\rm PDR}$/L$_{\rm TOT}$ ratio. They also tend to be associated with a higher fraction of 
more evolved UV emitting stars, as shown by their higher FUV-to-H$\alpha$ luminosity ratios compared to the NOSUB sample. In contrast, the single 
peak sources of the NOSUB sample are likely more isolated young star--forming regions.

Next we compare the subsamples of sources that either demonstrate a molecular gas detection (as traced by CO), with a signal-to-noise 
ratio higher than three times the rms-noise level, or demonstrate less significant or no detection. As defined in Section \S6, we refer to these subsamples as HighCO and LowCO respectively.
A similar KS-test shows that the two subsamples present different distributions for the dust temperature, total dust luminosity,   
the dust-corrected H$\alpha$ luminosity, the fraction of obscured H$\alpha$ luminosity and the gas column density.

The HighCO sample has warmer dust temperatures (median values of 22.3 K) than the LowCO sample (median values of 18.8 K).   Since the masses of both samples are comparable, this translates into higher dust luminosities for the HighCO sample (median dust luminosity of 4.0 $\times$ 10$^{39}$ erg s$^{-1}$) and lower dust luminosities for the LowCO sample (median value of 1.6 $\times$ 10$^{39}$ erg s$^{-1}$).     We also note that the HighCO sample show higher dust-corrected H$\alpha$ luminosities with median values twice as high as  the LowCO sample. The relative amount of  obscured H$\alpha$ luminosity to total dust H$\alpha$ luminosity is higher for the HighCO sample with a 
median value of 0.3 in comparison to a median value 0.2 for the LowCO sample.  As expected, the HighCO samples show higher gas column densities with a median value of 24 M$_\odot$ pc$^{-2}$ in comparison to 12 M$_\odot$ pc$^{-2}$ for the LowCO sample.

The differences seen between the HighCO and the LowCO sample suggest that the HighCO sample is characterized by a higher star formation 
activity. This is suggested by the higher H$\alpha$ dust-corrected luminosity in the HighCO sample, while the distribution of dust 
mass is quite similar to that of the LowCO sample. The star formation in the HighCO sample is also plausibly producing a 
stronger radiation field, which leads the dust temperature and the dust luminosity to higher values than the LowCO sample. 
This is to be expected given that star formation is tightly coupled to molecular gas (i.e. Bigiel et al.\ 2008).

Finally, we note that whether the sources display a HighCO or LowCO detection show no obvious relation to the H$\alpha$ 
morphology. About 68\% of the entire 
source sample is classified as NOSUB. Within the HighCO sample, 73\% of the sources are from the NOSUB sample, while the fraction reduces 
to 62\% in the LowCO sample. Thus, the NOSUB sources are only slightly more represented in the HighCO sample and slightly underrepresented 
in the LowCO sample.

 \begin{table}
\begin{center}
\caption{$p$-values  from KS-tests}
\begin{tabular}{ccc}
\hline

Property & SUB-NOSUB & HighCO-LowCO \\
\hline
\hline
T$_{\rm SED}$ & 0.24 & $<$ 0.01 \\

M$_{\rm SED}$ & 0.076 & 0.39 \\

L$_{\rm TOT}$ & 0.10 & $<$0.01 \\

L(H$\alpha$, cor) & 0.54 & $<$0.01 \\

L(FUV,cor)/ L(H$\alpha$, cor) & 0.016 & 0.32 \\

L$_{\rm PDR}/$L$_{\rm TOT}$ & 0.013 & 0.62 \\

L(H$\alpha$, obscured)/L(H$\alpha$, cor) & 0.20 & $<$0.01 \\

$\Sigma_{\rm gas}$ & 0.48 & $<$0.01 \\
\hline

\end{tabular}
\end{center}
\label{KS_tests}
\end{table}

\subsection{Radial Trends}

\begin{figure*}
\centering
\includegraphics[trim=0mm 40mm 0mm 0mm,width=150mm,angle=-90]{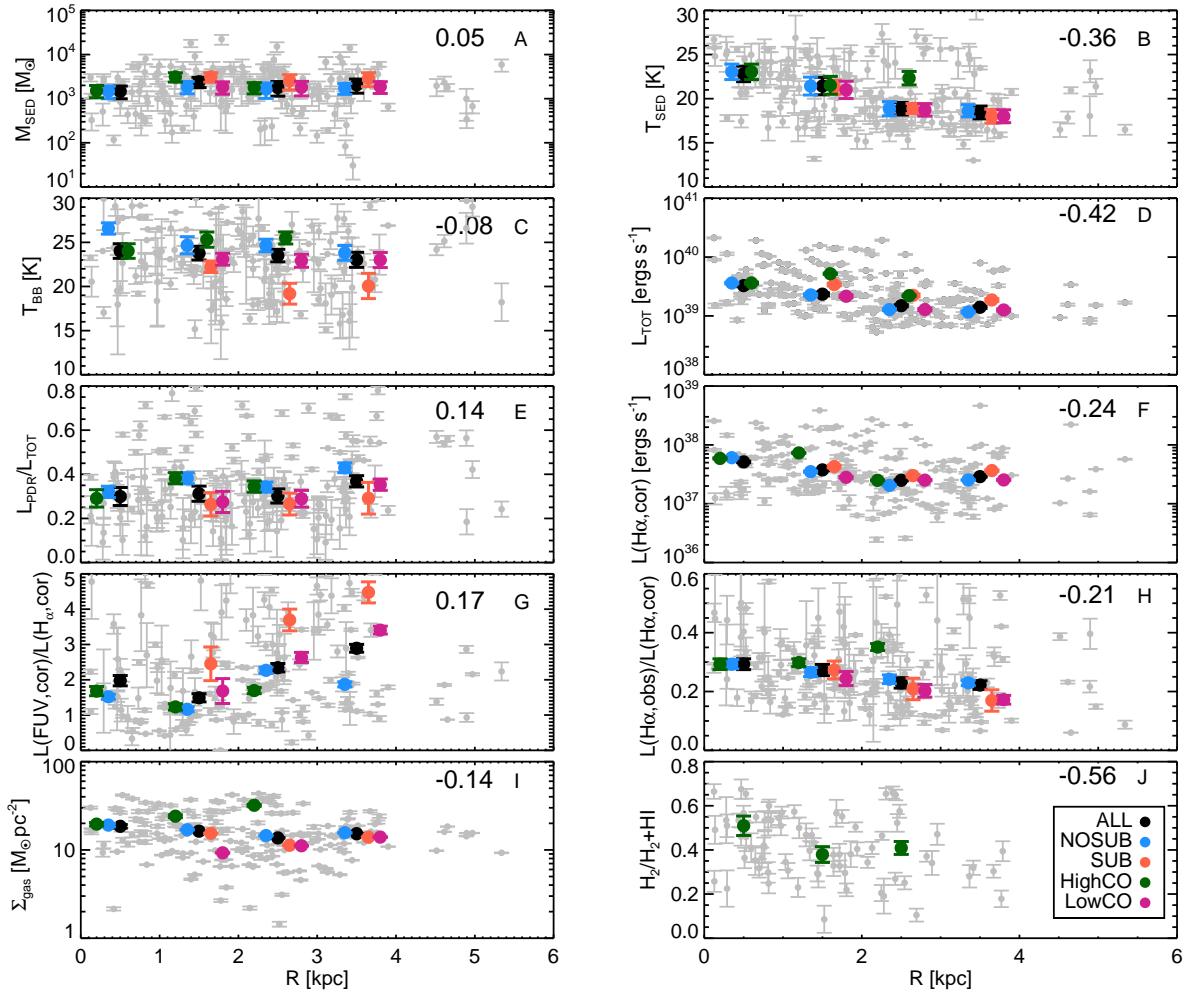}

\caption{Source parameters plotted versus deprojected radial distance. (a) Dust mass; (b) two component SED fit temperature T$_{\rm SED}$, 
(c) modified blackbody dust temperature T$_{\rm BB}$, (d) total dust luminosity, (e) PDR--to--total dust emission luminosity ratio, (f) 
dust corrected H$\alpha$ luminosity, (g) dust corrected FUV to H$\alpha$ luminosity ratio, (h) obscured to total H$\alpha$ luminosity 
ratio, (i) total gas mass surface density within the 350$\mu$m source footprint, (j) fraction of molecular gas (only for HighCO sources).
Each source is displayed in grey and the Spearman rank correlation coefficient for the entire sample is listed in the upper right.  
The median values in bins of 1 kpc are displayed in colour according to the legend in panel (j) and are slightly offset for visual 
clarity.  The uncertainties on the coloured points reflect median uncertainties.}
\label{radial_plots}
\end{figure*}

In Fig.~\ref{radial_plots} we examine how the inferred properties of the FIR detected sources in M33 vary 
with radius. We assume an inclination of 56$^{o}$ and a position angle of 23$^{o}$ (Corbelli \& Schneider 1997) in order to 
determine the deprojected radial position. We show the values for the entire sample (grey points) and also show the median values in radial bins of 1 kpc width for the four different subsamples. 

We find that, in general, the properties of the compact sources do not show strong radial dependencies (see the Spearman rank correlation 
coefficients in top right of each panel), although some parameters present mild or weak radial trends.   
While the source dust mass do not vary much with radius (panel A), the dust temperature of the sources, 
as measured by the two--component SED fit, decreases with radius and is on average 20\% higher inside 2 kpc (panel B). 
Interestingly, the dust temperature inferred by the single component 
modified blackbody fit (panel C) does not show a radial variation$^1$\footnote{$^1$The fact that the two measured dust temperatures do not 
show the same trend is not inconsistent since the modified blackbody temperature is 
obtained by fitting the observed fluxes over the entire FIR range, while the dust temperature from the two--component SED 
fit refers only to the diffuse dust emission component (see Sect. 4)}. We also find that the dust 
luminosity shows a mild negative trend with radius (panel D) and is, on average, at least a factor of two greater within 2 kpc 
compared to larger radii. The PDR--to--total dust luminosity ratio presents only a very weak positive trend with radius (panel E). 
A modest declining trend is shown by both the dust corrected H$\alpha$ luminosity (panel F) and the fraction of obscured 
H$\alpha$ luminosity (panel H, see also Sect. 7.3.2 for discussion on this result).  
Instead the FUV--to--H$\alpha$ luminosity ratio shows only a weak increase with radius (panel G). 
 
Unlike the source dust mass, the gas column density in the regions associated with the sources shows a small decline with 
radius, with sources in the inner 2 kpc having a column density 25\% higher (panel I). 
We note that LowCO sources are mainly found at radii higher than 1 kpc while HighCO sources are rare beyond 3 kpc.   
Therefore, the radial trend shown by the gas column density for the entire sample probably reflects the radial decline 
of the fraction of sources well--detected in CO (note also that, taken individually, both LowCO and HighCO sources seem actually
to show an increasing radial trend in panel I). 
Finally, the relative fraction of molecular gas for the HighCO subsample appears to decrease with radius, 
with the sources within the inner 2 kpc having more than double molecular gas fraction than the sources at larger radii (panel J). 
We note, however, that a negative metallicity radial gradient in the galaxy would act to increase the amount of molecular gas 
relative to the measured CO flux for the sources at larger radii (Galametz et al. 2011).  However, recent studies have shown that the metallicity gradient in M33 is shallow,
with a central abundance of 12 + log(O/H) =8.36 and a gradient of -0.027 dex kpc$^{-1}$ (Rosolowsky \& Simon 2008).

The change in molecular gas fraction could be linked to the other observed trends, namely for the dust luminosity and the diffuse 
dust temperature as well as the weaker trends for the H$\alpha$ luminosity and the fraction of obscured H$\alpha$ luminosity. 
All these trends are consistent with the sources presenting a higher amount of star formation at shorter radii as well as being more 
opaque to the radiation emitted by the local young stellar populations. Further insights on the relation between local star formation 
and dust heating in the sources will be given in the next section. 

\subsection{Dust heating within the FIR sources}

\subsubsection{Dust luminosity and H$\alpha$ luminosity}
\begin{figure}
\centering
\includegraphics[trim=0mm 0mm 0mm 0mm,width=80mm,angle=0]{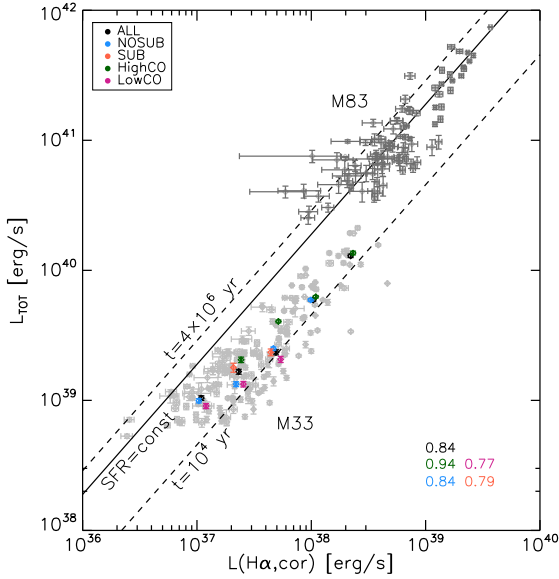}
\caption{Dust luminosity versus H$\alpha$ line luminosity for the FIR detected sources in M33 (this work) and in M83 (see FN13). 
The continuous line represents the total stellar population luminosity for a given amount 
of H$\alpha$ luminosity in the case of a constant SFR. The dashed lines correspond to the stellar luminosity in the case of a single--age 
stellar population with ages $t=10^4$\,yr and $t=10^6$\,yr. The median values of the sources within bins of size 0.32 dex are displayed 
using the same colour scheme as Fig.~\ref{radial_plots}.  The correlation coefficient between the dust and H$\alpha$ luminosity is 
listed for each sample in the bottom right.}
\label{Ldust_vs_lhalpha}
\end{figure}

The complete source sample as well as all the subsamples show a strong correlation between the total 
dust luminosity L$_{\rm TOT}$ and the dust corrected H$\alpha$ luminosity L(H$\alpha$,cor) (see Fig.\ref{Ldust_vs_lhalpha}). 
As discussed in FN13, the presence of 
this correlation is expected if  the dust luminosity is predominantly heated by the local young stellar populations provided one can assume
that the luminosity-weighted optical depth $\tau$ (integrated over wavelength) is similar for all the sources. 
Interestingly, for a given H$\alpha$ luminosity, the LowCO sources have dust luminosities which are systematically lower than 
the HighCO sources. This could be another indication that the HighCO sources are more obscured star formation regions, 
which are expected to be more opaque and thus absorbing radiation from the embedded young stars more efficiently. 
Fig.\ref{Ldust_vs_lhalpha} shows no clear differences between the SUB and NOSUB sources.  However, the SUB sources are confined to 
a narrower range of values than the NOSUB sources. We note that in Fig.\ref{Ldust_vs_lhalpha} median values are not plotted if there are less than 10 sources in a bin. 

If one assumes a continuous star formation history for the FIR detected sources, one can derive the amount of total luminosity emitted 
by the cospatial young stellar populations for a given H$\alpha$ luminosity. To this purpose, we consider the 
SFR-L$_{\rm dust}$ relation from Calzetti (2012), which assumes the IMF from Kroupa et al. (2001) 
and solar metallicity:
\begin{equation}
 {\rm SFR[M}_\odot{\rm /yr]=2.8}\times10^{-44}{\rm L}_{\rm dust} {\rm [erg/s].}
\end{equation}
In this equation, L$_{\rm dust}$ coincides with the total young stellar luminosity while the SFR can be considered in this case directly 
proportional to the H$\alpha$ luminosity, that is, SFR[M$_\odot$/yr]=5.3$\times$10$^{-42}$L(H$\alpha$,cor)[erg/s] (Calzetti et al. 2007). 
By combining the two previous relations, one can obtain the total young stellar luminosity as a function of L(H$\alpha$,cor).  
This relation is plotted in Fig.\ref{Ldust_vs_lhalpha} as a solid line and denotes the maximum dust luminosity that can be powered by 
the local young stellar populations assuming a constant SFR. We notice that the line lies well above the inferred source 
dust luminosities for the M33 sources for each value of L(H$\alpha$,cor). This shows that 
the total intrinsic luminosity of the young stellar populations is more than sufficient to power the observed dust emission in the 
assumption of a constant SFR. 

One can also compare the observed data with the theoretical prediction for the total young stellar luminosity  
in the case of an instantaneous burst of star formation (that is, a single--age stellar population). 
In this case the ratio between the bolometric stellar luminosity and the H$\alpha$ luminosity is not constant but increases with time
(since the H$\alpha$ emission fades away quicker than the stellar luminosity). We used \textsc{Starburst99} to derive the predicted 
 ratio of the total stellar 
luminosity to the H$\alpha$ luminosity for an instantaneous burst of star formation as a function of time. In this calculation we 
assume the standard input parameters of \textsc{Starburst99} including the same IMF and metallicity as before. The 
dashed lines in Fig.\ref{Ldust_vs_lhalpha} represent the total stellar luminosity of a young stellar population for each value of 
H$\alpha$ at times $t$=10$^4$\,yr and $t$=4$\times$10$^6$\,yr from the initial event of star formation. Similarly as before, 
these lines denote the maximum dust luminosity that can be powered by the local young stellar populations at the two selected ages.  
We note that most of the observed points in M33 lie between 
these two curves. This means that, assuming that the star formation regions are older than few $10^6$\,yr, the amount of total stellar 
luminosity is enough to power the observed dust emission in this case as well. Thus, with reasonable assumptions on the star formation 
history for these sources, it seems that the young stellar population luminosity is sufficient to account for the entire dust emission.

This finding can be compared with the analogous result we obtained in FN13 for the FIR bright sources in M83.  In FN13, however, 
we found evidence for extra--heating in addition to that provided by the local young stellar populations. 
For ease of comparison, we plot the sources detected in M83 in  
Fig.\ref{Ldust_vs_lhalpha}, using the so called ``High background measurements'' in that paper. From this figure, one can see 
that the sources in M83 are more luminous in both dust and H$\alpha$ emission. This is likely due to the fact that 
 the spatial resolution in M83 is lower (130\, pc at 70$\mu$m versus 28\, pc at 100$\mu$m for M33). Thus, the detected compact sources are 
 more massive and presumably contain multiple embedded star--forming regions. However, more importantly, it is also evident that the 
L$_{\rm TOT}$-L(H$\alpha$,cor) 
ratio for the M83
sample is higher than that observed for the M33 sample. This is seen by comparing the position of the points for the 
two source samples with the line corresponding to a constant SFR. As suggested in FN13, additional heating from the 
local old stellar populations or the diffuse ISRF of the galaxy may boost the dust luminosity close to or above 
the constant SFR line. This additional dust heating for the M83 sources
could approximately compensate the fraction of young stellar luminosity escaping unabsorbed from each source. 
In M33, however, the heating from local young stellar populations seems 
enough to account for the entire dust emission. We note that the higher dust to H$\alpha$ luminosity ratio for the M83 
sources 
might also be related to the different spatial resolution of two studies. 
As a 
consequence, the sources detected in M83 might contain a larger fraction of cold dust, plausibly located in between the multiple star 
formation regions within each FIR source. This dust component might be heated significantly by older or non-local stellar population 
apart from the local young stars.

\subsubsection{Dust-to-H$\alpha$ luminosity ratio versus radial distance}

\begin{figure}
\centering
\includegraphics[trim=0mm 0mm 0mm 0mm,width=80mm,angle=0]{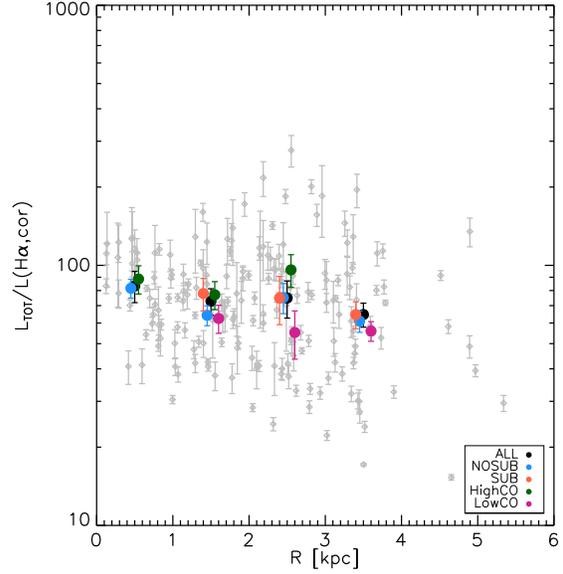}
\caption{Ratio of the dust luminosity to the H$\alpha$ luminosity versus the deprojected radial distance for the full sample 
(grey points).  Median values in bins of 1 kpc size are shown for each subsample.}
\label{ldust_sfr_ratio_vs_rad}
\end{figure}

We also examine whether the ISRF due to the total stellar population of M33 is a significant source of 
dust heating, powering at least a fraction of the dust emission from the compact sources. This can be the case if the molecular 
clouds associated with the FIR sources are also heated externally by the galactic ISRF.  To do so we plot the 
ratio of the dust-to-H$\alpha$ luminosity versus the deprojected radial distance. In fact, if both 
the ISRF and local star formation play a 
significant role in heating the dust, the source dust-to-H$\alpha$ luminosity ratio can be written as: 
\begin{equation}
\frac{L_{\rm TOT}}{L(\rm{H\alpha,cor})}=\frac{L_{\rm TOT}^{\rm ISRF}(r)}{L(\rm{H\alpha,cor})}+\frac{L_{\rm TOT}^{\rm loc}}{L(\rm{H\alpha,cor})}
\end{equation}
where L$_{\rm TOT}^{\rm ISRF}$ is the dust luminosity powered by the galaxy ISRF and L$_{\rm TOT}^{\rm loc}$
is the dust luminosity due to the absorption of local young stellar luminosity. We expect this ratio to decline with radius if 
 1) L$_{\rm TOT}^{\rm ISRF}$ is not negligible and declines with radius following the decline of the galaxy interstellar 
 radiation field; 
 2) the H$\alpha$ luminosity does not vary systematically with radius or it varies differently than L$_{\rm TOT}^{\rm ISRF}$;  
 3) the ratio L$_{\rm TOT}^{\rm loc}$/L(H$\alpha$,cor), proportional to the source local opacity, is on average similar at all radii. 
 As shown in Fig.\ref{ldust_sfr_ratio_vs_rad}, there is at most only a modest declining trend for this ratio 
with radial distance (the Spearman's rank correlation coefficient for the entire sample is equal to -0.19). 
This result is consistent with only a small fraction of the total dust emission heated by the galaxy ISRF. 
However, as shown in Sect. 7.2, the absorbed fraction of H$\alpha$ luminosity shows a small decline with radius. One might expect 
 this quantity to be at least roughly proportional to L$_{\rm TOT}^{\rm loc}$/L(H$\alpha$,cor). Therefore, it might be that the small 
 radial trend observed for $L_{\rm TOT}/L(\rm{H\alpha,cor})$ is due to the fact that, although the sources are heated 
 predominantly by the local star formation, at shorter radii 
 they are simply obscuring more efficiently the radiation from the local young stellar populations. If this is the case, the 
 term $L_{\rm TOT}^{\rm ISRF}(r)$ would not actually be needed to justify the observed small radial trend for $L_{\rm TOT}/L(\rm{H\alpha,cor})$. 
 This interpretation is consistent with a negligible contribution from the galaxy ISRF to the compact source 
 dust heating.

\subsubsection{Dust mass -- dust temperature anticorrelation}
\begin{figure}
\centering
\includegraphics[trim=0mm 0mm 0mm 0mm,width=80mm,angle=0]{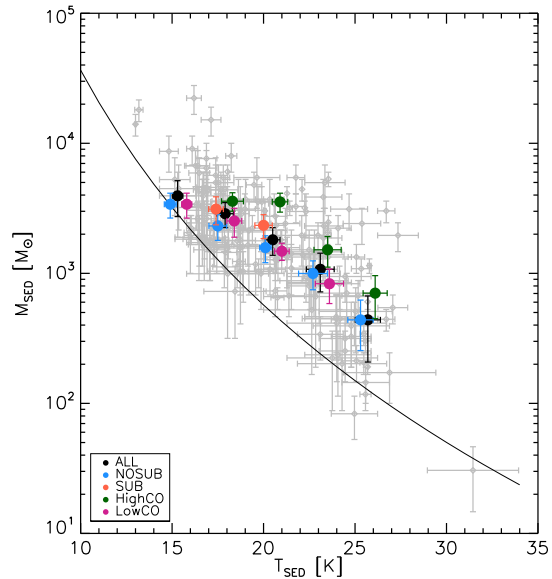}
\caption{Inferred dust mass versus dust temperature from the MIR/FIR SED fit for the entire sample (grey points).  
We show median values within bins of size 2.6 K for each of the subsamples. The plotted curve pass through points 
of same modified blackbody luminosity (see the text for details).}
\label{Mdust_vs_Tdust}
\end{figure}

An anticorrelation between dust mass and dust temperature has been observed by FN13 for the sample of FIR compact sources 
in M83. FN13 suggested that 
the lack of points in the lower left part of the dust mass - dust temperature plot was due to the lower limit for the dust luminosity of 
the sources detectable on the FIR maps of M83. For M33 we find a similar effect. Plotting the dust mass versus dust temperature 
for the entire source sample (see 
Fig.\ref{Mdust_vs_Tdust}), we find that 
there is an anticorrelation between the two parameters which is also present for the single subsamples separately. 
This result is reminiscent of the dust temperature--dust spectral index (T-$\beta$) anticorrelation found when fitting modified blackbody curves to FIR data of 
nearby galaxies (e.g. Galametz et al. 2012).  
While the T-$\beta$ anticorrelation may be due to an intrinsic variation of dust properties in the ISM of galaxies, it is also possible that this anticorrelation is due to the effect of 
data noise combined with the intrinsic degeneracy of the T and $\beta$ parameters in determining the spectral 
shape of emission within limited wavelength ranges (see Shetty et al. 2009). 

However, at variance with the T and $\beta$ anticorrelation, there is no intrinsic degeneracy between T$_{\rm SED}$ and M$_{\rm SED}$, which would produce similar spectral shapes
for different pairs of these parameters and thus contribute to the observed anticorrelation between these two.  While a higher value of T$_{\rm SED}$ requires a lower M$_{\rm SED}$  
to fit a certain monochromatic flux, we note that the spectral shape of the templates used in the source SED fit is affected only by T$_{\rm SED}$ but not by M$_{\rm SED}$,
which is only a scaling factor. 

Instead, as in the case of M83, there are several 
indications that the observed trend is mostly due to the 
lower limit of the observed dust luminosity for the well-detected sources. 
In order to show this, we plot in Fig. \ref{Mdust_vs_Tdust} a curve such that $M_{{\rm dust}}T^{6}_{{\rm dust}}$ is constant. The points belonging 
to this curve represent the $M_{{\rm dust}}$ and $T_{{\rm dust}}$ combinations which give the same total luminosity
for a modified blackbody spectra $k_\lambda B_\lambda(T_{{\rm dust}}$) with $k_\lambda\propto \lambda^{-2}$. Specifically, the corresponding luminosity 
for the plotted curve is $\approx 8\times10^{38}$\,erg/s and we note that the curve is roughly consistent with the lower boundary for the region occupied 
by the measurements$^2$\footnote{$^2$Note that, although our SED templates cannot be approximated 
by a single modified blackbody, the latter reproduces the FIR bump typically containing most of the dust luminosity.}.  However, this lower limit for the dust luminosity is determined by how well the 
fluxes in the single infrared wavebands are measured. That is, if the lower boundary in the M$_{\rm SED}$--T$_{\rm SED}$ plane is due to the flux detection
limits, one should find that, for a given combination of M$_{\rm SED}$ and T$_{\rm SED}$ below the curve, the predicted monochromatic fluxes are below
the detection thresholds. We checked this by predicting the observed fluxes in the FIR, using our theoretical diffuse dust emission 
templates used in the two component SED fitting procedure, assuming combinations of dust mass and temperature in that area. The predicted fluxes are 
either well below or, for combinations of dust mass and temperature close to the solid curve in Fig.\ref{Mdust_vs_Tdust}, 
just comparable with those of the faintest sources we detected with \textsc{getsources} (after the removal of sources with low detection 
significance, see \S6). For example for T$_{\rm SED}$=20~K 
and M$_{\rm SED}$=4$\times$10$^2$~M$_\odot$ (just below the plotted line), the predicted 
100$\mu m$ flux is about 0.08 Jy which is at the lower end of the flux distribution of our source sample. 

We also note a rather scarcity of sources in the upper right part of the M$_{\rm SED}$-T$_{\rm SED}$ diagram. The lack of a homogeneously 
distributed high mass--high temperature sources could be due 
to physical effects at variance with the lack of sources in the lower left part of the diagram. Alternatively, more massive clouds 
might simply appear fragmented when observed in dust emission, and thus detected as multiple less massive sources. 

When we consider the median trends for the different samples separately, there are no particular differences except the fact that the 
LowCO sources have systematically lower median dust masses compared to the HighCO ones for a given dust temperature. 
 
We conclude that the observed anticorrelation is mainly due to the lower limit in source luminosity in our sample, causing the lower 
boundary of the region where the sources are confined, and perhaps a different physical (or observational) effect could 
be responsible for determining a lower density of points in the upper-right part of the diagram. It is important to take into account the existence 
of this correlation when quantities depending on 
dust mass and temperature are plotted versus each other. Observed correlations might be misinterpreted if one overlooks the relation
between these two parameters.  

\subsection{Gas column densities, GMC dust masses and star formation}
\begin{figure}
\centering
\includegraphics[trim=0mm 0mm 0mm 0mm,width=100mm,angle=-90]{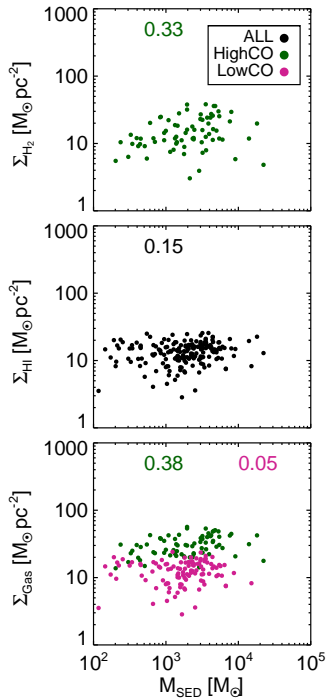}
\caption{Molecular, atomic and total gas mass surface density versus source dust mass.  
The correlation coefficients are shown at the top of each panel according to the coding in the legend.}
\label{dust_gas}
\end{figure}

We now examine how the source dust masses compare to the local gas column densities in that region.
 We make note here that the average column density 
of the gas has been measured using aperture photometry with fairly large apertures that scale with the size of the source at 350$\mu$m.  
In contrast, the source dust mass has been determined by extracting the bright FIR sources and subtracting the background emission.  
While dust is generally well-mixed with gas, we would not necessarily anticipate that the mass of the clumpy source would scale with 
average column density in a given region.  In Fig.~\ref{dust_gas} we plot the local average column density of the gas versus the dust mass of the clumpy sources. In general, we do not find a strong correlation 
between the source dust mass and the average gas column density. In addition to the total gas column density we also  consider the gas phases and plot the molecular gas column density (HighCO sources only) and the atomic gas column density for all sources. 
The different gas phases also do not show evident correlations.  
This suggests that  a wide range of inferred GMC masses can be found in regions of similar gas density (averaged over ~100pc).

\begin{figure*}
\centering
\includegraphics[trim=0mm 30mm 0mm 30mm,width=80mm,angle=-90]{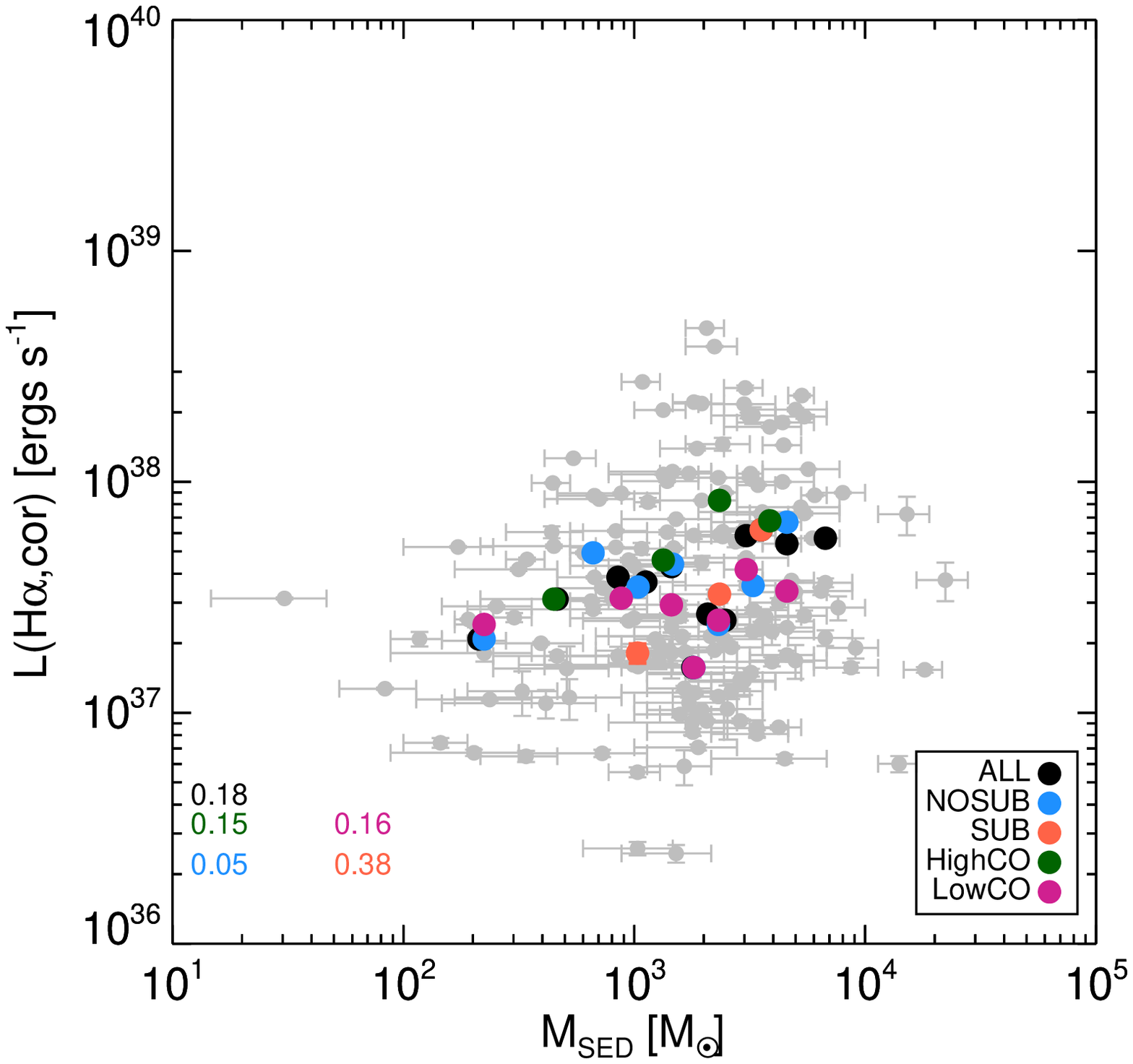}
\includegraphics[trim=0mm 30mm 0mm 30mm,width=80mm,angle=-90]{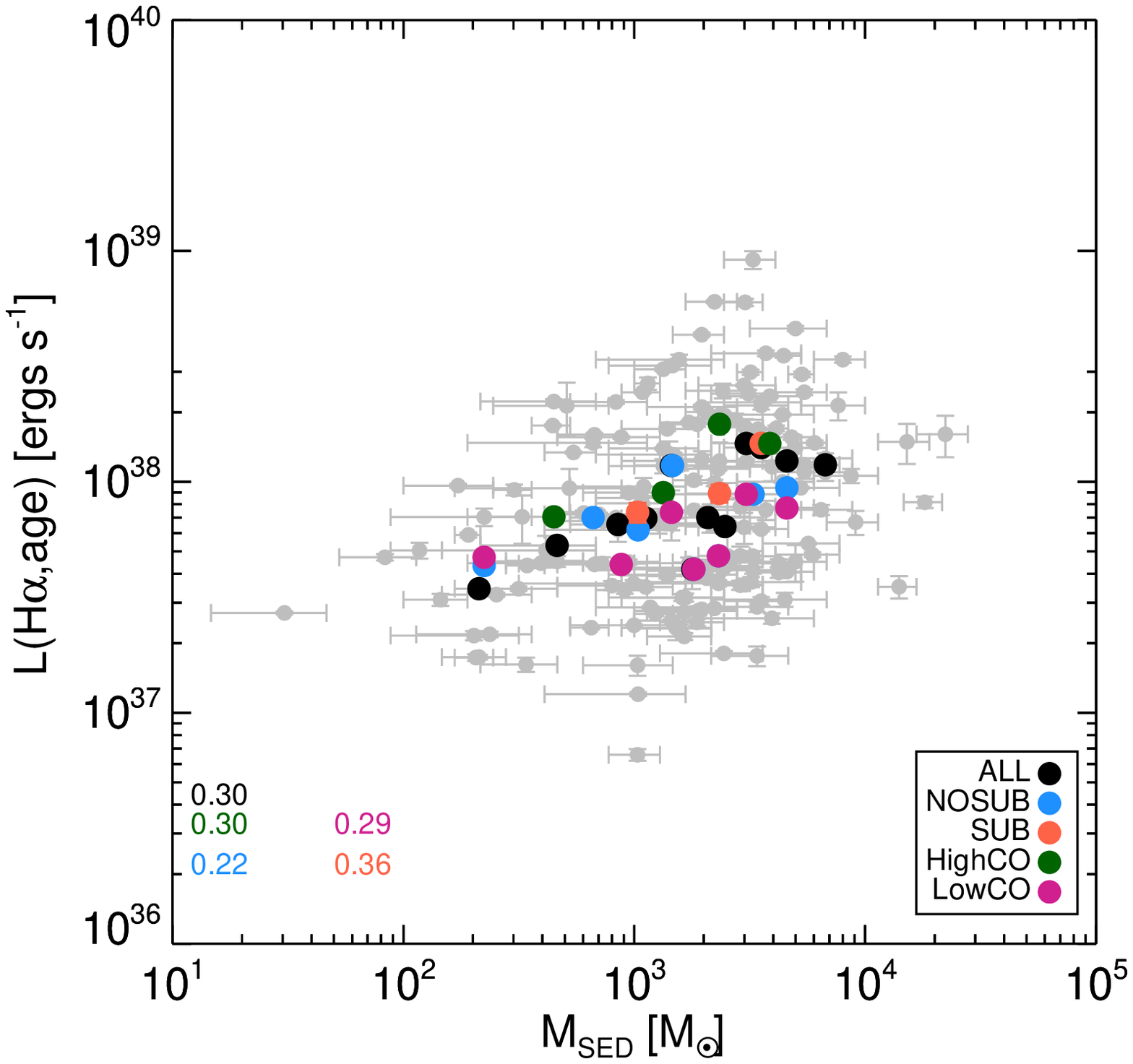}
\caption{Source dust-corrected H$\alpha$ luminosity versus dust mass (left) and H$\alpha$ luminosity corrected for both dust attenuation 
and age effects versus dust mass (right).  The source measurements are plotted in grey and median values for each subsample are plotted 
according to the legend.  The median values are determined in five equidistant bins of dust mass with at least 10 sources in each bin. 
The correlation coefficients for the complete set of points in each subsample are noted on the bottom left of each panel.}  
\label{sf_mass}
\end{figure*}

We also examine how the total dust-corrected H$\alpha$ luminosity 
scales with source mass (left-hand panel of Fig.~\ref{sf_mass}). If we assume that the dust traces gas and that the dust-corrected 
H$\alpha$ luminosity is tracing recent star formation, 
this plot is equivalent to plotting the total star formation rate versus 
the gas mass of the GMCs associated with the FIR sources$^3$\footnotetext[1]{$^3$Note that the H$\alpha$ 
luminosity is proportional to the SFR only for a constant star formation history within about 10$^7$\,yr.}.  

On large scales it is well-known that the star formation rate per unit area is well-correlated to the gas column density 
(Kennicutt--Schmidt relation; Kennicutt 1989).  
While Fig.~\ref{sf_mass} is similar there are 
important differences: 1) we are plotting total values instead of surface brightnesses; 2) the source sizes are much 
smaller than the typical regions used to recover such a relation. Indeed, we do not recover a strong correlation between the mass of 
the GMCs and the dust-corrected H$\alpha$ luminosity.  This is not surprising particularly given the scales we are probing 
(see e.g. Onodera et al.\ 2010 and Schruba et al. 2010). 

One of the plausible reasons for the breakdown of the correlation is the spread in ages of the young stellar populations 
associated with the different clouds, which causes the intrinsic H$\alpha$ luminosity to vary given the same amount of young 
stellar mass. Therefore, one might expect that if we can 
correct the H$\alpha$ luminosity for the stellar population age, we could recover the relation or at least reduce the observed scatter. 
We attempt to do this in the right-hand panel of Fig.~\ref{sf_mass}.  Here we plot the dust-corrected H$\alpha$ luminosity that has been 
corrected for the source age, assuming that each source can be modelled by an instantaneous 
starburst, as described in \S5. We find that the correlation coefficients increase and that the scatter is reduced.  
This result suggests that the spread in ages of the stellar populations in the different sources is indeed one of the important 
factors determining the breakdown of observed star formation relations at the scales of GMCs. These results will be further 
discussed in Sect. 8.2.


\section{Discussion}

\subsection{The origin of the clumpy FIR emission in M33}
In this work we sought to determine the physical origin of a sample of clumpy FIR sources detected on the {\it Herschel} 
maps of M33. This requires an understanding of both the kind of ISM clouds where the emitting dust is located and the radiation sources 
heating the dust. About the former, the main indication comes from the measurement of the source dust mass. As shown in \S7, we found that 
the FIR sources we analysed in this paper are associated with dust masses in the range 
10$^2$-10$^4$ M$_\odot$. By assuming a typical gas-to-dust mass ratio equal to 100, this implies gas masses between 
10$^4$ and 10$^6$ M$_\odot$ which are in the range of GMC masses. This result suggests that our technique can be considered as an 
alternative way to study GMCs in nearby external galaxies, compared to more typical gas line emission studies (e.g. 
Wilson \& Scoville 1990, Engargiola et al. 2003, Rosolowsky et al. 2003 and 2007, Bigiel et al. 2010, Gratier et al. 2010).
In this work, we found that for sources of similar dust mass (and, thus, plausibly similar gas mass) it is not obvious that they 
will be both detected in CO emission equally well. This finding led us to separate the sources into the HighCO and LowCO sub-samples and 
we showed that on average these samples present some different characteristics but still similar dust masses. The variation in CO 
brightness for clouds of similar dust mass might be due to intrinsic differences in both the relative amount of CO gas inside a cloud 
and the CO line emissivity due to local gas conditions and ISRF. If this is the case, using 
dust emission to detect GMCs and measure the total amount of gas within them could be considered as a more robust way than using CO
line emission. However, one should mention that using this approach can also be affected by a number of caveats (apart from the typical 
uncertainties on the assumed dust optical properties and gas-to-dust mass ratio).  
In fact, it is not obvious that whole GMCs, associated with one or more young stellar clusters,  can be clearly detectable 
in dust emission. The appearance of a dusty GMC as a compact source on a FIR galaxy map depends on several factors such as 
(1) how well the GMC dust is illuminated by the embedded star formation, the local old stellar populations and from the diffuse 
ISRF, (2) how massive the GMC is (because $L_{\rm dust}$ $\propto$ $M_{\rm dust }T^{6}_{\rm dust}$, sources 
with low masses will not be easily detected), (3) how variable is the background FIR emission of the galaxy considered.   
In addition, in the hypothesis that a substantial part of the dust heating is due to embedded star formation, GMCs might not be 
entirely detectable on FIR maps for a series of physical reasons: (1) star formation is in the very early stages and not enough young 
stellar emission is produced to make the GMC distinguishable from the FIR background emission (especially at shorter 
FIR bands such as 70--160$\mu$m); (2) the embedded stellar population is efficiently heating only a fraction of the GMC while the rest is 
not distinguishable from the background; (3) the stellar populations are producing winds and shocks which are separating the gas and 
dust from the stars. In this way, the dust-stars geometry becomes more complex and gradually the GMC dust emission fades into the 
background emission.         

The FIR emission from the sources seems to be powered mainly by local star 
formation. The FIR sources are cospatial 
with H$\alpha$/UV emission and the extrapolated young stellar population 
luminosity is enough to power the observed dust emission luminosity (a conclusion based on reasonable assumptions on the star formation 
history, see Sect. 7.3). 
A good correlation between the total dust luminosity L$_{\rm TOT}$ and the dust-corrected H$\alpha$ luminosity L$(\rm{H\alpha,cor})$ is 
observed, which would be expected if (1) the dust heating is
 due to recent embedded star formation and (2) the relative fraction of radiation from young stellar populations escaping unabsorbed from 
 the clouds is similar for all the sources (see also section 2.2 of KN13). Internal heating due to star formation is also consistent with 
 the lack of a strong radial decline of L$_{\rm TOT}$/L$(H\alpha,cor)$, which might be
expected when the diffuse ISRF provides a significant contribution to the dust heating (see Sect. 7.3.2). 

These results can be compared with the findings of two previous works containing an analysis of the dust emission 
from bright HII regions in M33, namely Komugi et al. (2011, see their section 5) and Hippelein et al. (2003, section 4). 
Komugi et al. (2011) showed that the cold dust 
temperature (estimated from the $160\mu m/1100 \mu m$ flux ratio) is 
better correlated with the $K_S$ band (2.1 $\mu m$) luminosity than with the dust-corrected H$\alpha$ luminosity for a sample 
of HII regions in M33. This result is used to claim that the cold dust is heated by local stellar populations that are older than those 
responsible for the H$\alpha$ emission. However, we note that those authors do not subtract a local background when performing the source 
photometry. Especially in the FIR, the background emission can be a large fraction of the total emission and physically is plausibly 
associated with a more diffuse dust component compared to that associated with star--forming GMCs. 

At variance with Komugi et al. (2011), while performing the source photometry, we subtract a local background at all wavelengths as 
described in \S3. Thus, we emphasize that in this work we claim that the \textit{background--subtracted} dust emission, 
observed at the positions of the FIR sources, is powered by the local young stellar populations. 
The inclusion of the background 
emission might well contribute to the correlation between K$_S$ band luminosity and cold dust temperature, found by Kumugi et al. (2011), 
 because of their mutual dependence on radius. Furthermore, we point out that
one should be careful when interpreting correlations between stellar population luminosity and 
dust temperature (see also section 2.2 of FN13). Given a certain amount of stellar luminosity injected locally in the ISM, the average 
dust temperature depends both on the dust mass being illuminated and the relative distribution of dust and stars (apart from the dust
optical properties). Also, especially 
when modelling the dust emission associated with diffuse galaxy ISM, radiative 
transfer effects should be taken into account to model properly the radiation field produced by 
non local stellar populations (see e.g. Popescu et al. 2011, Baes et al. 2011).

For these reasons, in order to investigate dust heating in the FIR sources, we preferred to show a comparison between the
luminosity of a local stellar population (namely, the young stars traced by the H$\alpha$ luminosity) and the dust luminosity. 
In case the local stellar population considered 
is responsible for heating the dust, these two quantities are expected to correlate provided that the relative fraction of locally 
absorbed stellar luminosity is similar for all the sources. In addition, as we discussed in \S7.3, in this way one is able to 
prove that the local stellar luminosity is high enough to power the dust emission. This cannot be done by comparing stellar luminosities 
with dust temperatures.

The weak decreasing radial trend we found for L$_{\rm TOT}$/L$(H\alpha,cor)$ in section 7.3.2 can be qualitatively compared with figure 7 of
Hippelein et al. (2003), where they plotted the H$\alpha$ to 60$\mu$m flux ratio versus radius for about 60 star-forming regions in M33
and found that this ratio increases significantly with the distance from the galaxy centre. They attributed this 
effect to decreasing H$\alpha$ attenuation towards large radii. Apart from the data and specific method used for the photometry, 
there are several major differences in comparison to our work: (1) Hippelein et al. (2003) plotted the source H$\alpha$ flux not corrected 
for dust attenuation; (2) they used the 60$\mu$m flux, a tracer of warm dust emission, while we consider the total wavelength integrated
dust luminosity; (3) they did not perform local background subtraction during the source photometry but they attempted to isolate the 
emission from the star--forming regions by assuming that there are fixed FIR colours for the dust emission from star formation regions and 
from the background component. Nonetheless, as discussed in 
section 7.3.2, it is plausible that the weak radial trend we found for L$_{\rm TOT}$/L$(H\alpha,cor)$ is indeed due to decreasing 
attenuation of the H$\alpha$ emission with radial distance. This interpretation is consistent with that provided by 
Hippelein et al. (2003) to explain their results.

To summarize, our analysis is consistent with the sources being GMCs heated by embedded star formation. This is reminiscent of 
the results of Verley et al. (2010), who found a correlation between H$\alpha$ and 250$\mu$m luminosity for a sample of 250$\mu$m detected
sources, and of Boquien et al. (2011), 
who found that, at the positions with the highest values of H$\alpha$ brightness on the disc of M33, the warm dust temperature is 
dictated by the H$\alpha$ brightness itself. 

Even though the FIR sources all seem to be star--forming GMCs, there are some differences both in the spatial 
distribution of star--forming regions within them (e.g. a single compact star--forming region versus multiple regions) and in their 
evolutionary stage which are evidenced by the morphology of H$\alpha$ emission, the dust corrected UV/H$\alpha$ luminosity ratio as 
well as the presence or absence of well detected CO emission. These differences are reflected by the average values of the dust 
parameters for the subsamples we created. In the following subsection, we discuss the results related to the H$\alpha$ morphological classification 
NOSUB and SUB while those related to the CO detection classification, HighCO and LowCO, will be discussed in the next section which 
deals with gas and star formation.

\subsubsection{Insights from the H$\alpha$ morphology of the FIR sources}

Compared to NOSUB sources, SUB sources are in particular characterized by a lower fraction 
of PDR luminosity contributing to the total dust emission and by higher values of the dust corrected UV/H$\alpha$ luminosity ratio. 
Together with their irregular 
morphology in H$\alpha$, these evidences suggest that SUB sources tend to be either sources composed by multiple star formation sites 
detected collectively in the FIR, or, in same cases, by a single star formation site in an evolved stage. Plausibly the diffuse dust 
located between multiple star formation sites contributes to increase the cold dust emission compared to the PDR contribution, 
thus the lower fraction of PDR luminosity observed for these sources. This is also expected for the sources in a more evolved stage, 
where the dust emission gradually becomes colder because dust is farer away from the central source (R13 found a similar
effect). Also, the fact that the UV/H$\alpha$ luminosity ratio is typically larger for SUB sources compared to NOSUB sources indicates 
that there is a higher fraction of older stellar populations in the SUB sample. We note that there are no apparent 
differences in the spatial location on the disc between SUB and NOSUB sources.  

From the point of view of the H$\alpha$ based morphological classification of the sources, the most similar analysis to the one we 
performed is the work by R13, who however selected their source sample based 
on the detection in H$\alpha$ not in the FIR. They selected a sample of HII regions in M33 and classified them depending on the
morphology in four types: “filled regions” (compact knot of emission), “mixed regions” (presenting several knots of emission 
and filamentary structures), “shells” (showing arcs in the form of a shell), “clear shells” (complete and close shells”). 
In our classification NOSUB sources resemble the “filled regions” and SUB sources ``the mixed regions''. However, we did not 
find clear evidence of ``shells'' in our sample. ``Shell'' sources in the R13 sample reside predominantly at large radii, since 
it is easier to identify this kind of sources in less crowded regions. The location of these sources is often outside the common
field of view of our data set, within which we limited our analysis. In addition, R13 found that the “mixed regions and filled regions” 
are the brightest in the FIR, while the surface brightness of ``shell'' sources is on average a factor $\approx$0.5-0.8 dex lower 
(see the right-hand panel of their fig. 5). Since we considered only sources with good detections in the FIR, this is consistent with ``shell'' objects 
not being well represented in our source sample.

\subsection{Gas and star formation properties of FIR sources}

In general, we find that presence of molecular gas is central to defining the properties of the FIR sources.  In particular, we find that the samples of HighCO and LowCO show differing dust luminosities, with the HighCO sources having higher dust luminosities.  We have also found that, on average, 
sources associated with molecular gas have higher H$\alpha$ luminosities and higher dust temperatures.  

Given that molecular gas is coupled with star formation (e.g. Bigiel et al. 2008), these changes in the dust properties are likely associated with the presence of increased star formation.  Higher dust temperatures of clouds in the molecular phase have also been found by 
Paradis et al. (2011) in the case of the LMC using modified blackbody fits.  These increased temperatures have been attributed to local heating.

The presence of molecular gas also appears to be responsible for the trends seen with radius.  The molecular gas fraction of the sources are highest in 
the inner regions.  
Studies of GMCs in M33 have also shown that there is a decrease in CO luminosity with radius (Gratier et al.\ 2010).  
In general, we find that HighCO sources are clustered more towards the centre and LowCO sources towards the outer parts 
(see Fig.~\ref{radial_plots}).  We find that sources show a mild radial decline in dust luminosity, opacity, and temperature  despite 
comparable masses.

While the radial decline in the molecular gas fraction is the most likely reason for these changes, we note that there are several other 
possibilities for these trends.  If the dust is heated also by a non-local population, the lower intensities of the interstellar 
radiation field in the outer regions could drive a decline in dust luminosity and temperature.  However, as noted in the previous section and in Sect. 7.3, it appears that sources are mostly locally heated.  
Secondly, the dust optical properties might change with radius.  
As found by Tabatabaei et al. (2014), there is some evidence that the dust emissivity changes with radius in M33.  
In our SED fitting we have assumed that the dust size and chemical composition do not change much throughout the galaxy.  However, if these properties were to change, this could affect the inferred dust temperature.  
However, Tabatabaei et al. (2014) also found that the differences could be attributed to  whether a region was star-forming or not.  That is, the emissivity was dependent on whether the region showed active star formation or was more diffuse.
Given that we have isolated only clumpy structures in the disc which are associated with star-forming regions and we are not considering 
diffuse regions, we do not expect the emissivity to vary.   In contrast, 
in Tabatabaei et al. (2014) azimuthal averages were employed to examine the radial decline of the dust emissivity.

While many FIR sources are classified as HighCO sources, over half of them do not have an associated molecular gas component above three times
the noise rms on the CO map (i.e. classified as LowCO sources).   If we assume that the FIR emission is powered by embedded SF regions, 
we ask why there are some sources without a CO detection. 
The most likely reason for this is that LowCO sources are more evolved and there has been gas dissipation.  In support of this,  we have seen that  LowCO sources have higher UV-to-H$\alpha$ ratios, which suggests more evolved clouds.  We also find that HighCO sources have a higher fraction of obscured H$\alpha$ luminosity, which suggests that HighCO sources may be in an earlier evolutionary phase.  However, the possibility also exists that in the LowCO regions we are not detecting molecular gas that may be present.  This may be H$_{2}$ that is not traced by CO or there is CO that we do not detect due to sensitivity issues.  In the GMC study of 
Engargiola et al. (2003) they find similar results.  As they note, there are many more HII regions than are GMCs.  They conclude that winds and ionizing radiation in more evolved star--forming regions may have dissociated the molecular clouds.  This conclusion is consistent with what we find.

In general, we find that  the average gas column density is not well-correlated with the total dust mass of the FIR sources. We note that the average gas column density in the region of each source was measured for both atomic and molecular gas using aperture photometry and provides a measure of the local average conditions and does not probe the GMC mass. 
Meanwhile, the dust mass of the sources is determined by an extraction process that considers the shape and size of the source and includes a local background subtraction.  In this way, provided that the dust mass can be translated into a gas mass by a multiplicative constant, we can trace the GMC mass using the dust mass of the sources.  Thus, we can examine whether the local environment 
plays a role in determining the mass of the sources detected.  The lack of a correlation between the local gas column density and 
the dust mass of sources suggests that a wide range of GMC masses can be found in regions with the same gas column density. 
This is supported by the fact 
that the source dust masses do not vary with radius, yet azimuthal averages of the total gas density have shown that it decreases with radius 
(Tosaki et al. 2011).  In a study of GMC properties in M33, Engargiola et al.\ 2003 also found that the cloud mass was independent of the local ambient HI mass.  

However, we note that it is possible that higher gas column densities could lead to 
an increased number of clouds.  Unfortunately, we cannot address this question because our sample is not complete. As explained in \S 6, we have cut a number of sources 
due to low signal-to-noise and the $\chi^{2}$ criteria of our SED fit. Thus, it is not possible to do proper statistics on the number of 
sources in certain locations.

We find only a weak correlation between dust-corrected H$\alpha$ luminosity and the dust mass of the sources.  If we can translate the dust mass directly into a gas mass and use the H$\alpha$ 
luminosity as a tracer of the star formation rates of the sources, we can explore the relation between star formation rates and the GMC masses.  We note that most studies use surface densities for these quantities 
(typically called a Kennicutt--Schmidt plot) 
and we use total values here.  As has been shown in many previous studies (e.g. Schruba et al. 2010 and Onodera et al. 2010) which have measured GMC masses directly, 
we do not find a strong correlation between the inferred GMC masses and our H$\alpha$ tracer of star formation.  
The so-called  ``breakdown of the KS relation'' at sub-kiloparsec scales in M33 has been attributed to the stochasticity of the star 
forming regions and the fact that the small regions do not provide a great enough stellar population to average over 
evolutionary states (e.g. Schruba et al. 2010 and Onodera et al. 2010).  
Given this, a possible way forward would be to correct the H$\alpha$ luminosity for age effects using stellar population models, 
assuming that the observed young stellar populations have been formed in an instantaneous starburst. In this case, 
the H$\alpha$ luminosity is not a tracer of the SFR but is rather proportional to the young stellar population
mass if one corrects it for the dimming due to the stellar population age. We attempted to do this using the ratio of the 
dust-corrected UV-to-H$\alpha$ luminosity (see Sect. 5 and 7.4).  
After applying the age correction, we found that the correlation between the source dust (or gas) mass and the H$\alpha$ luminosity improved and the scatter 
between the two was reduced. This hints that a more thorough approach incorporating a detailed analysis of stellar emission SEDs
 may be a necessary step for the determination of a star formation law at GMC scales. 

Apart from the different ages of the stellar populations within the clouds, other physical factors can lead to increased scatter in the H$\alpha$ luminosity -- dust mass diagram. 
In particular, different GMCs might be characterized by 
a different gas density distribution, determined by the cloud interstellar turbulence and/or magnetic fields 
(e.g. Federrath \& Klessen 2012).  
This can lead to a spread in star formation efficiency for a given cloud mass, since only the densest regions within a GMC are able to 
collapse and form stars. In addition, the fraction of molecular gas in the clouds and, thus, the fraction of the cloud mass able 
to form stars is limited by the cloud's ability to be shielded by the UV ISRF (e.g. Krumholz et al. 2009). 
In the end, it is likely a combination of all or some of these factors, which can be of both internal and external origin, that define 
the observed variations of GMC star formation efficiency (see Dobbs et al. 2013 for a recent review on the topic).

\section{Summary and conclusions}

In this work we have presented a study of the dust, star formation and gas properties of a sample of FIR 
compact sources in M33. We performed the photometry of the sources both on the FIR maps from {\it Herschel}
and on a set of multiwavelength data, including MIR, H$\alpha$, FUV, as well as integrated HI and CO gas line maps. We fit 
the dust emission SED in order to measure the dust mass, temperature and luminosity parameters and we estimated the dust 
corrected H$\alpha$ and FUV fluxes, by using the 24$\mu$m emission as 
a proxy for dust attenuation. We also used the dust-corrected FUV/H$\alpha$ ratio in order to estimate the age of the stellar 
populations associated with the FIR sources and we corrected the H$\alpha$ luminosity for age effects (assuming an instantaneous 
starburst).
In order to differentiate between sources of intrinsically different types, we categorized the sources based on whether or not they 
demonstrated substructured in H$\alpha$ (SUB and NOSUB) and whether or not they demonstrated  a significant  ($>$3$\sigma$) 
CO detection within the source footprint (HighCO and LowCO).

The main results of this work are the following:\\ 
(i) The sources have dust masses in the range $10^2$--$10^4M_\odot$, dust temperatures between 15-35K and dust luminosities of
$10^{39} -10^{40}$ erg s$^{-1}$;\\
(ii) Compared to NOSUB sources, SUB sources are characterized by a lower PDR-to-total dust luminosity ratio and higher UV-to-H$\alpha$ 
luminosity ratio;\\
(iii) Compared with LowCO sources, HighCO sources are characterized by higher dust luminosities, higher dust corrected H$\alpha$ 
luminosities, lower UV-to-H$\alpha$ luminosity ratios and higher fractions of obscured H$\alpha$ luminosity;\\
(iv) We find that the source properties do not show strong trends with radial distance. However, the total dust luminosity 
and the cold dust temperature tend to decrease mildly with radius. Weaker trends are also shown by the dust corrected H$\alpha$ 
luminosity and by the fraction of absorbed H$\alpha$ luminosity. The HighCO sources in the inner regions are also embedded 
in an environment with a higher molecular gas fraction.  \\
(v) The source dust luminosities and dust-corrected H$\alpha$ luminosities are well correlated. The bolometric young stellar population 
luminosity, extrapolated from the H$\alpha$ luminosity, assuming both a constant SFR or a single--age stellar population 
with ages of few times $10^6$\,yr, is well above the observed dust luminosity. Furthermore, the dust-to-H$\alpha$ luminosity 
present only a weak trend with radial distance. All these findings are consistent with dust being heated predominantly by local 
star formation; \\
(vi) We observe a dust mass -- dust temperature anticorrelation, which is at least partially due to the lower limit to the dust luminosity 
of the detectable sources;\\
(vii) We did not find a clear trend between the source dust masses and the local gas column density; \\
(viii) We did not find a good correlation between the dust--corrected H$\alpha$ luminosity and the dust mass of the sources. However, the 
scatter is substantially reduced by adopting the age--corrected H$\alpha$ luminosity.

\section*{Acknowledgements}
We thank an anonymous referee for useful suggestions that helped
us to improve the paper. GN acknowledges support from the UK 
Science and Technology Facilities Council (STFC; grant
ST/J001341/1, P.I. C. C. Popescu) and thanks the Max Planck Institute f\"{u}r 
Kernphysik for support during the completion of this work. KF thanks P. Gratier for helpful conversations and acknowledges  
support by grants from the Canadian Space Agency and the Natural Science and Engineering Research Council of Canada (PI: C. D. Wilson).

\appendix

\section{Source Photometry}

\onecolumn
\begin{center}
\tiny
\begin{longtable}{ccccccccccccc}
\caption{Source positions and photometry.}

\label{source_fluxes}\\

\hline
 & RA & $\delta$ & F$_{8\mu m}$ & F$_{24\mu m}$ & F$_{100\mu m}$ & F$_{160\mu m}$ & F$_{250\mu m}$ & F$_{350\mu m}$ & F$(H_\alpha)$ & F(FUV) & $<\Sigma_{HI}>$ & $<\Sigma_{H2}>$\\
 & [deg] & [deg]& [mJy] & [mJy] & [mJy] & [mJy] & [mJy] & [mJy] & [10$^{-14}$ erg/s/cm$^{2}$]    & [mJy] & [M$_\odot$/pc$^2$] & [M$_\odot$/pc$^2$] \\
\hline
\endfirsthead
\multicolumn{3}{c}%
{\tablename\ \thetable{} -- continued from previous page}\\
\hline
 & RA & $\delta$ & F$_{8\mu m}$ & F$_{24\mu m}$ & F$_{100\mu m}$ & F$_{160\mu m}$ & F$_{250\mu m}$ & F$_{350\mu m}$ & F$(H_\alpha)$ & F(FUV) & $<\Sigma_{HI}>$ & $<\Sigma_{H2}>$ \\
 & [deg] & [deg]& [mJy] & [mJy] & [mJy] & [mJy] & [mJy] & [mJy] & [10$^{-14}$ erg/s/cm$^{2}$]  & [mJy]  & [M$_\odot$/pc$^2$] & [M$_\odot$/pc$^2$] \\
\hline
\endhead

       1 &        23.556225 &        30.562494 & 92 $\pm$ 4.9 & 164 $\pm$ 7.2 & 
2750 $\pm$ 140 & 2500 $\pm$ 210&1420 $\pm$ 210&760 $\pm$ 130&99 $\pm$ 2.6&
11.71 $\pm$ 0.58&37&45\\
       2 &        23.397340 &        30.658550 & 51 $\pm$ 2.7 & 149 $\pm$ 6.5 & 
1810 $\pm$ 100 & 1850 $\pm$ 150&1230 $\pm$ 160&620 $\pm$ 100&43 $\pm$ 1.4&
3.78 $\pm$ 0.18&24&52\\
       3 &        23.295836 &        30.754879 & 45 $\pm$ 2.3 & 152 $\pm$ 6.6 & 
2270 $\pm$ 120 & 1790 $\pm$ 140&860 $\pm$ 120&559 $\pm$ 81&186 $\pm$ 3.9&
20.90 $\pm$ 0.90&29&--\\
       4 &        23.438708 &        30.613953 & 53 $\pm$ 3.2 & 99 $\pm$ 4.6 & 
1385 $\pm$ 82 & 1240 $\pm$ 110&590 $\pm$ 120&330 $\pm$ 97&55 $\pm$ 1.5&
5.82 $\pm$ 0.26&22&28\\
       5 &        23.374106 &        30.530245 & 155 $\pm$ 8.5 & 222 $\pm$ 11 & 
4020 $\pm$ 230 & 2900 $\pm$ 280&1020 $\pm$ 190&463 $\pm$ 99&160.5 $\pm$ 4.0&
15 $\pm$ 2.1&41&56\\
       6 &        23.434809 &        30.743958 & 74 $\pm$ 5.4 & 177 $\pm$ 8.3 & 
2490 $\pm$ 130 & 1780 $\pm$ 140&900 $\pm$ 140&334 $\pm$ 74&163 $\pm$ 3.9&
14.2 $\pm$ 2.0&16&21\\
       7 &        23.493667 &        30.812199 & 35 $\pm$ 1.8 & 57 $\pm$ 2.5 & 
1119 $\pm$ 65 & 1330 $\pm$ 100&790 $\pm$ 140&470 $\pm$ 80&20.21 $\pm$ 0.51&
5.69 $\pm$ 0.21&23&63\\
       8 &        23.394513 &        30.618688 & 40 $\pm$ 3.7 & 101 $\pm$ 5.4 & 
1479 $\pm$ 87 & 1080 $\pm$ 130&720 $\pm$ 160&343 $\pm$ 94&74.5 $\pm$ 2.0&
3.78 $\pm$ 0.17&30&22\\
\end{longtable}
\end{center}

\begin{center}
\tiny
\begin{longtable}{ccccccccc}
\caption{Source SED best fitting parameters and source classification.}\\
\label{source_sed_table}\\

\hline
 & $\chi_{\text{UV}}$ & $\chi_{\text {col}}$ & T$_{\text {SED}}$ & M$_{\text {SED}}$ & F$_{24}$ & L$_{\text {TOT}}$ & $\chi^2_{\text{FIT}}$ & TYPE \\
 &                   &                     &    [K]   &       [$10^2~M_\odot$]  &     &   [$10^{38}$erg/s]  &   \\     
\hline
 \endfirsthead
\multicolumn{3}{c}%
{\tablename\ \thetable{} -- continued from previous page}\\
\hline
& $\chi_{\text{UV}}$ & $\chi_{\text {col}}$ & T$_{\text {SED}}$ & M$_{\text {SED}}$ & F$_{24}$ & L$_{\text {TOT}}$ & $\chi^2_{\text{FIT}}$ & TYPE \\
 &                   &                     &    [K]   &       [$10^2~M_\odot$]  &     &   [$10^{38}$erg/s]  &   \\    
\hline
\endhead

       1 & 2$\pm$1 & 2$\pm$1 & 22.1$\pm$0.7 & 44.3$\pm$8.4 & 0.79$\pm$0.10
 & 136.0$\pm$4.2 &   4.6&NOSUB - HighCO\\
       2 & 0.5$\pm$0.2 & 4$\pm$2 & 18.2$\pm$0.3 & 80$\pm$20 & 0.92$\pm$0.12 & 
103.0$\pm$2.2 &   1.3&NOSUB - HighCO\\
       3 & 1.3$\pm$0.8 & 9$\pm$2 & 22$\pm$1 & 30$\pm$11 & 0.93$\pm$0.12 & 110.6
$\pm$2.2 &   5.3&NOSUB - LowCO\\
       4 & 3$\pm$2 & 1.2$\pm$0.9 & 22$\pm$1 & 19.6$\pm$8.2 & 0.88$\pm$0.17 & 
74.2$\pm$2.3 &   1.4&NOSUB - HighCO\\
       5 & 9$\pm$2 & 2$\pm$1 & 27$\pm$1 & 19.6$\pm$4.9 & 0.68$\pm$0.13 & 193.6
$\pm$5.4 &   1.4&NOSUB - HighCO\\
       6 & 5$\pm$2 & 3$\pm$2 & 25.1$\pm$0.4 & 13.3$\pm$3.3 & 0.87$\pm$0.11 & 
131.1$\pm$1.7 &   0.8&NOSUB - HighCO\\
       7 & 0.7$\pm$0.2 & 3.$\pm$1 & 19.5$\pm$0.4 & 48$\pm$12 & 0.789$\pm$0.10
 & 52.92$\pm$0.46 &   1.1&NOSUB - HighCO\\
       8 & 3$\pm$2 & 3$\pm$2 & 23$\pm$1 & 13.9$\pm$5.1 & 0.89$\pm$0.11 & 75.47
$\pm$0.85 &   2.6&NOSUB - HighCO\\
\end{longtable}
\end{center}


\clearpage

\begin{thebibliography}{}

\bibitem[Aniano et al.(2011)]{2011PASP..123.1218A} Aniano, G., Draine, B.~T., Gordon, K.~D., \& Sandstrom, K.\ 2011, \pasp, 123, 1218 
\bibitem[Barker \& Sarajedini(2008)]{2008MNRAS.390..863B} Barker, M.~K., \& Sarajedini, A.\ 2008, \mnras, 390, 863 
\bibitem[Baes et al. (2011)]{2011ApJS..196...22B} Baes, M., Verstappen, J. , De Looze, I.,  Fritz, J. , Saftly, W. 2011, ApJS, 196, 22
\bibitem[Bendo et al. (2012)]{2012MNRAS.419.1833B}  Bendo, G. J., Boselli, A., Dariush, A., Pohlen, M., Roussel, H. et al.  2012, MNRAS, 419, 1833
\bibitem[Bigiel et al. (2008)]{2008AJ....136.2846B} Bigiel, F., Leroy, A., Walter, F., Brinks, E. et al. 2008, AJ, 136, 2846
\bibitem[Bigiel et al. (2010)]{2010ApJ...725.1159B} Bigiel, F., Bolatto, A. D., Leroy, A. K., Blitz, L., Walter, F. et al. 2010, ApJ, 725, 1159
\bibitem[Boquien et al. (2011)]{2011AJ....142..111B} Boquien, M., Calzetti, D., Combes, F., Henkel, C., Israel, F., Kramer, C. et al.  2011, AJ, 142, 111
\bibitem[Braine et al. (2010)]{2010A&A...518L..69B} Braine, J., Gratier, P., Kramer, C., Xilouris, E. M., Rosolowsky, E. et al. 2010, A\&A, 518, 69
\bibitem[Calzetti et al. (2007)]{2007ApJ...666..870C} Calzetti, D., Kennicutt, R. C., Engelbracht, C. W., Leitherer, C., Draine, B. T., et al. 2007, ApJ, 666, 870
\bibitem[Calzetti(2012)]{2012arXiv1208.2997C} Calzetti, D.\ 2012, arXiv:1208.2997 
\bibitem[Calzetti et al.(2012)]{2012ApJ...752...98C} Calzetti, D., Liu, G., \& Koda, J.\ 2012, \apj, 752, 98
\bibitem[Corbelli \& Schneider (1997)]{1997ApJ...479..244C} Corbelli, E., Schneider, S. E. 1997, ApJ, 479, 244
\bibitem[Dale et al.(2009)]{2009ApJ...703..517D} Dale, D.~A., Cohen, S.~A., Johnson, L.~C., et al.\ 2009, \apj, 703, 517 
\bibitem[Dobbs et al. (2013)]{2013arXiv1312.3223D} Dobbs, C. L., Krumholz, M, R., Ballesteros-Paredes, J., Bolatto, A. D. et al. 2013arXiv1312.3223D
\bibitem[Engargiola et al. (2003)]{2003ApJS..149..343E} Engargiola, G., Plambeck, R. L., Rosolowsky, E., Blitz, L. 2003, ApJS, 149, 343
\bibitem[Federrath \& Klessen (2012)]{2012ApJ...761..156F} Federrath, C., Klessen, R. S. 2012, ApJ, 761, 156
\bibitem[Foyle et al. (2012)]{2012MNRAS.421.2917F} Foyle, K., Wilson, C. D., Mentuch, E., Bendo, G., Dariush, A., et al. \ 2012, MNRAS, 421, 2917
\bibitem[Foyle et al. (2013)]{2013MNRAS.432.2182F} Foyle, K., Natale, G., Wilson, C. D., Popescu, C. C., Baes, M. et al. 2013, MNRAS, 432, 2182	
\bibitem[Galametz et al. (2011)]{2011A&A...532A..56G} Galametz, M., Madden, S. C., Galliano, F., Hony, S. et al. 2011, A\&A, 532, 56
\bibitem[Galametz et al. (2012)]{2012MNRAS.425..763G} 	Galametz, M., Kennicutt, R. C., Albrecht, M., Aniano, G., Armus, L. et al. 2012, MNRAS, 425, 763
\bibitem[Gil de Paz et al. (2007)]{2007ApJS..173..185G} Gil de Paz, A. et al. 2007, ApJS, 173, 185 
\bibitem[Gratier et al. (2010)]{2010A&A...522A...3G} Gratier, P., Braine, J., Rodriguez-Fernandez, N. J., Schuster, K. F., Kramer, C. et al. 2010, A\&A, 522, 3
\bibitem[Greenawalt (1998)]{1998PhDT........16G} Greenawalt, B. E. Thesis (PHD), NEW MEXICO STATE UNIVERSITY, 1998
\bibitem[Griffin et al.(2010)]{2010A&A...518L...3G} Griffin, M.~J., et al.\ 2010, A\&A, 518, L3 
\bibitem[Helou et al. (2004)]{2004ApJS..154..253H} Helou, G., Roussel, H., Appleton, P., Frayer, D. et al. 2004, ApJS, 154, 253
\bibitem[Hippelein et al. (2003)]{2003A&A...407..137H} Hippelein, H., Haas, M., Tuffs, R. J., Lemke, D., Stickel, M. et al. 	2003, A\&A, 407, 137
\bibitem[Hoopes \& Walterbros (2000)]{2000ApJ...541..597H} Hoopes, C. G., Walterbos, R. A. M. 2000, ApJ, 541, 597
\bibitem[Hughes et al. (2013)]{2013ApJ...779...46H} Hughes, A., Meidt, S. E., Colombo, D., Schinnerer, E. et al. 2013, ApJ, 779, 46
\bibitem[Kennicutt(1989)]{1989ApJ...344..685K} Kennicutt, R.~C., Jr.\ 1989, \apj, 344, 685
\bibitem[Kennicutt(1998)]{1998ApJ...498..541K} Kennicutt, R.~C., Jr.\ 1998, \apj, 498, 541 
\bibitem[Kennicutt et al.(2007)]{2007ApJ...671..333K} Kennicutt, R.~C., Jr., Calzetti, D., Walter, F., et al.\ 2007, \apj, 671, 333
\bibitem[Kennicutt \& Evans (2012)]{2012ARA&A..50..531K} Kennicutt, Robert C., Evans, Neal J. 2012, ARA\&A, 50, 531K
\bibitem[Komugi et al. (2011)]{2011PASJ...63.1139K} Komugi, S., Tosaki, T., Kohno, K., Tsukagoshi, T. et al. 2011, PASJ, 63, 1139
\bibitem[Kramer et al. (2010)]{2010A&A...518L..67K} Kramer, C., Buchbender, C., Xilouris, E. M., Boquien, M., Braine, J. et al. 2010, A\&A, 518, 67
\bibitem[Kramer et al. (2013)]{2013A&A...553A.114K} Kramer, C., Abreu-Vicente, J., García-Burillo, S., Relaño, M., Aalto, S. et al. 2013, A\&A, 553, 114
\bibitem[Kroupa (2001)]{2001MNRAS.322..231K} Kroupa, P. 2001, MNRAS, 322, 231
\bibitem[Krumholz et al. (2009)]{2009ApJ...699..850K} 	Krumholz, M. R., McKee, C, F., Tumlinson, J. 2009, ApJ, 699, 850
\bibitem[Leitherer et al. (1999)]{1999ApJS..123....3L} Leitherer, C., Schaerer, D., Goldader, J. D., Gonzalez Delgado, R. M. et al. 1999, ApJS, 123, 3
\bibitem[Leroy et al.(2008)]{2008AJ....136.2782L} Leroy, A.~K., Walter, F., Brinks, E., Bigiel, F., de Blok, W.~J.~G., Madore, B., \& Thornley, M.~D.\ 2008, AJ, 136, 2782 
\bibitem[Martin et al. (2005)]{2005ApJ...619L...1M} Martin, D. C. et al. 2005, ApJ, 619, 1
\bibitem[Mathis et al. (1983)]{1983A&A...128..212M} Mathis, J. S., Mezger, P. G., Panagia, N. \ 1983, A\&A, 128, 212
\bibitem[Men'shchikov et al. (2012)]{2012A&A...542A..81M} Men'shchikov, A., Andr\'e, Ph., Didelon, P., Motte, F., Hennemann, M., Schneider, N. \ 2012, A\&A, 542, 81
\bibitem[Mentuch-Cooper et al. (2012)]{2012ApJ...755..165M} Mentuch Cooper, E., Wilson, C. D., Foyle, K., Bendo, G., Koda, J. et al. 2012, ApJ, 755, 165
\bibitem[Natale et al. (2010)]{2010ApJ...725..955N} Natale, G., Tuffs, R. J., Xu, C. K., Popescu, C. C., Fischera, J., Lisenfeld, U., et al. \ 2010, \apj, 725, 955
\bibitem[Onodera et al. (2010)]{2010ApJ...722L.127O} Onodera, S., Kuno, N., Tosaki, T., Kohno, K. et al. 2010, ApJ, 722, 127O
\bibitem[Paradis et al. (2011)][2011AJ....141...43P] Paradis, D. et al. 2011, AJ, 141, 43
\bibitem[Poglitsch et al.(2010)]{2010A&A...518L...2P} Poglitsch, A., et al.\ 2010, A\&A, 518, L2 
\bibitem[Popescu et al. (2011)]{2011A&A...527A.109P} Popescu, C. C., Tuffs, R. J., Dopita, M. A., Fischera, J., Kylafis, N. D., Madore, B. F. \ 2011, A\&A,527,109
\bibitem[Rela{\~n}o \& Kennicutt (2009)]{2009ApJ...699.1125R} Rela{\~n}o, M., Kennicutt, R. C. Jr. 2009, ApJ, 699, 1125
\bibitem[Rela{\~n}o et al.(2013)]{2013A&A...552A.140R} Rela{\~n}o, M., Verley, S., Pérez, I., Kramer, C., Calzetti, D. et al. 2013, A\&A, 552, 140
\bibitem[Rohlfs \& Wilson(1996)]{1996tra..book.....R} Rohlfs, K., \& Wilson, T.~L.\ 1996, Tools of Radio Astronomy, XVI, 423 pp.~127 figs., 20 tabs..~ Springer-Verlag Berlin Heidelberg New York.~Also Astronomy and Astrophysics Library,  
\bibitem[Rosolowsky \& Simon (2008)]{2008ApJ...675.1213R} Rosolowsky, E., Simon, Joshua D. 2008, ApJ, 675, 1213
\bibitem[Rosolowsky et al. (2003)]{2003ApJ...599..258R} Rosolowsky, E., Engargiola, G., Plambeck, R., Blitz, L. 2003, ApJ, 599, 258
\bibitem[Rosolowsky et al. (2007)]{2007ApJ...661..830R} Rosolowsky, E., Keto, E., Matsushita, S.,  Willner, S. P. 2007, ApJ, 661, 830
\bibitem[Schlegel et al.(1998)]{1998ApJ...500..525S} Schlegel, D.~J., Finkbeiner, D.~P., \& Davis, M.\ 1998, \apj, 500, 525
\bibitem[Schruba et al.(2010)]{2010ApJ...722.1699S} Schruba, A., Leroy, A.~K., Walter, F., Sandstrom, K., \& Rosolowsky, E.\ 2010, \apj, 722, 1699
\bibitem[Sharma et al. (2011)]{2011A&A...534A..96S} Sharma, S., Corbelli, E., Giovanardi, C., Hunt, L. K., Palla, F. 2011, A\&A, 534, 96
\bibitem[Shetty et al. (2009)]{2009ApJ...696..676S} Shetty, R., Kauffmann, J., Schnee, S., Goodman, A. A. 2009, ApJ, 696, 676
\bibitem[Smith et al.(2010)]{2010A&A...518L..51S} Smith, M. W. L., Vlahakis, C., Baes, M., Bendo, G. J., Bianchi, S. et al. 	2010, A\&A, 518, 51
\bibitem[Strong et al. (1988)]{1988A&A...207....1S} Strong, A. W., Bloemen, J. B. G. M., Dame, T. M., Grenier, I. A. et al. 	1988, A\&A, 207, 1
\bibitem[Tabatabaei et al. (2014)]{2014A&A...561A..95T} Tabatabaei, F. S., Braine, J., Xilouris, E. M., Kramer, C., Boquien, M. et al. 2014, A\&A, 561, 95
\bibitem[Thilker et al. (2005)]{2005ApJ...619L..67T} Thilker et al. 2005, ApJ, 619, 67
\bibitem[Tosaki et al. (2011)]{2011PASJ...63.1171T} Tosaki, T., Kuno, N., Onodera, Sachiko Miura, R. et al. 2011, PASJ, 63, 1171
\bibitem[Verley et al. (2010)]{2010A&A...518L..68V} Verley, S., Relaño, M., Kramer, C., Xilouris, E. M., Boquien, M., et al. 2010, A\&A, 518, 68
\bibitem[Xilouris et al. (2012)]{2012A&A...543A..74X} Xilouris, E. M., Tabatabaei, F. S., Boquien, M., Kramer, C., Buchbender, C. et al. 2012, A\&A, 543, 74
\bibitem[Wilson \& Scoville (1990)]{1990ApJ...363..435W} Wilson, C. D., Scoville, N. 1990, ApJ, 363, 435


\end{thebibliography}
\end{document}